\pdfoutput=1
\documentclass[12pt,a4paper]{article}

\usepackage{ifthen} 
\newboolean{pdflatex}
\setboolean{pdflatex}{true} 

\newboolean{articletitles}
\setboolean{articletitles}{true} 

\newboolean{uprightparticles}
\setboolean{uprightparticles}{false} 

\def\paperauthors{{Alessio Gianelle$^1$}, {Patrick Koppenburg$^2$}, {Donatella Lucchesi$^{1,3}$}, {Davide Nicotra$^{3,4}$}, {Eduardo Rodrigues$^5$}, {Lorenzo Sestini$^1$}, {Jacco de Vries$^4$}, {Davide Zuliani$^{1,3,6}$}} 

\def\paperasciititle{Quantum Machine Learning for $b$-jet charge identification} 
\def\papertitle{Quantum Machine Learning for $b$-jet charge identification} 
\def\paperkeywords{{High Energy Physics}, {LHCb}} 
\def\papercopyright{\the\year\ CERN for the benefit of the LHCb collaboration} 
\def\paperlicence{CC BY 4.0 licence}
\def\paperlicenceurl{https://creativecommons.org/licenses/by/4.0/}

\usepackage[top=1in, bottom=1.25in, left=1in, right=1in]{geometry}

\usepackage{microtype}
\usepackage{lineno}  
\usepackage{xspace} 
\usepackage{caption} 

\usepackage{graphicx}  
\usepackage{color}
\usepackage{colortbl}
\graphicspath{{./figs/}} 

\usepackage{amsmath} 
\usepackage{amssymb}
\usepackage{amsfonts}
\usepackage{upgreek} 

\newcommand*\patchAmsMathEnvironmentForLineno[1]{%
\expandafter\let\csname old#1\expandafter\endcsname\csname #1\endcsname
\expandafter\let\csname oldend#1\expandafter\endcsname\csname
end#1\endcsname
 \renewenvironment{#1}%
   {\linenomath\csname old#1\endcsname}%
   {\csname oldend#1\endcsname\endlinenomath}%
}
\newcommand*\patchBothAmsMathEnvironmentsForLineno[1]{%
  \patchAmsMathEnvironmentForLineno{#1}%
  \patchAmsMathEnvironmentForLineno{#1*}%
}
\AtBeginDocument{%
\patchBothAmsMathEnvironmentsForLineno{equation}%
\patchBothAmsMathEnvironmentsForLineno{align}%
\patchBothAmsMathEnvironmentsForLineno{flalign}%
\patchBothAmsMathEnvironmentsForLineno{alignat}%
\patchBothAmsMathEnvironmentsForLineno{gather}%
\patchBothAmsMathEnvironmentsForLineno{multline}%
\patchBothAmsMathEnvironmentsForLineno{eqnarray}%
}

\usepackage{hyperxmp}

\usepackage[pdftex,
            pdfauthor={\paperauthors},
            pdftitle={\paperasciititle},
            pdfkeywords={\paperkeywords},
            pdfcopyright={Copyright (C) \papercopyright},
            pdflicenseurl={\paperlicenceurl}]{hyperref}

\usepackage[colorinlistoftodos,textsize=scriptsize]{todonotes}

\usepackage[bottom,flushmargin,hang,multiple]{footmisc}

\usepackage[all]{hypcap} 

\usepackage{xspace} 
\usepackage{upgreek}

\def\MagUp {\mbox{\em Mag\kern -0.05em Up}\xspace}

\ifthenelse{\boolean{uprightparticles}}%
{

 \def\PDelta      {\ensuremath{\Delta}\xspace}                 
 \def\PXi         {\ensuremath{\Xi}\xspace}                 
 \def\PLambda     {\ensuremath{\Lambda}\xspace}                 
 \def\PSigma      {\ensuremath{\Sigma}\xspace}                 
 \def\POmega      {\ensuremath{\Omega}\xspace}                 
 \def\PUpsilon    {\ensuremath{\Upsilon}\xspace}

 \def\PB      {\ensuremath{\mathrm{B}}\xspace}                 
                  
 \def\PD      {\ensuremath{\mathrm{D}}\xspace}

 \def\PK      {\ensuremath{\mathrm{K}}\xspace}

 \def\Pi      {\ensuremath{\mathrm{i}}\xspace}

 \def\Ps      {\ensuremath{\mathrm{s}}\xspace}

 \def\thebaroffset{0.0em}
}
{

 \mathchardef\PDelta="7101
 \mathchardef\PXi="7104
 \mathchardef\PLambda="7103
 \mathchardef\PSigma="7106
 \mathchardef\POmega="710A
 \mathchardef\PUpsilon="7107
                  
 \def\PB      {\ensuremath{B}\xspace}                 
                  
 \def\PD      {\ensuremath{D}\xspace}

 \def\PK      {\ensuremath{K}\xspace}

 \def\Pi      {\ensuremath{i}\xspace}

 \def\Ps      {\ensuremath{s}\xspace}

 \def\thebaroffset{0.18em}
}
\newcommand{\offsetoverline}[2][\thebaroffset]{\kern #1\overline{\kern -#1 #2}}%

\makeatletter
\ifcase \@ptsize \relax
  \newcommand{\miniscule}{\@setfontsize\miniscule{4}{5}}
\or
  \newcommand{\miniscule}{\@setfontsize\miniscule{5}{6}}
\or
  \newcommand{\miniscule}{\@setfontsize\miniscule{5}{6}}
\fi
\makeatother

\DeclareRobustCommand{\optbar}[1]{\shortstack{{\miniscule (\rule[.5ex]{1.25em}{.18mm})}
  \\ [-.7ex] $#1$}}

\def\squark    {{\ensuremath{\Ps}}\xspace}

\def\KorKbar {\kern \thebaroffset\optbar{\kern -\thebaroffset \PK}{}\xspace}

\def\D       {{\ensuremath{\PD}}\xspace}

\def\DorDbar {\kern \thebaroffset\optbar{\kern -\thebaroffset \PD}\xspace}

\def\Dp      {{\ensuremath{\D^+}}\xspace}
\def\Dm      {{\ensuremath{\D^-}}\xspace}

\def\DpDm    {\ensuremath{\Dp {\kern -0.16em \Dm}}\xspace}

\def\B       {{\ensuremath{\PB}}\xspace}

\def\BorBbar {\kern \thebaroffset\optbar{\kern -\thebaroffset \PB}\xspace}

\def\Bd      {{\ensuremath{\B^0}}\xspace}

\def\BdorBdbar {\kern \thebaroffset\optbar{\kern -\thebaroffset \Bd}\xspace}

\def\Bs      {{\ensuremath{\B^0_\squark}}\xspace}

\def\BsorBsbar {\kern \thebaroffset\optbar{\kern -\thebaroffset \Bs}\xspace}

\def\Y#1S{\ensuremath{\PUpsilon{(#1S)}}\xspace}

\def\LorLbar     {\kern \thebaroffset\optbar{\kern -\thebaroffset \PLambda}\xspace}

\def\AT#1     {\ensuremath{A_{\mathrm{T}}^{#1}}\xspace}           

\def\C#1      {\ensuremath{\mathcal{C}_{#1}}\xspace}                       
\def\Cp#1     {\ensuremath{\mathcal{C}_{#1}^{'}}\xspace}                    
\def\Ceff#1   {\ensuremath{\mathcal{C}_{#1}^{\mathrm{(eff)}}}\xspace}        
\def\Cpeff#1  {\ensuremath{\mathcal{C}_{#1}^{'\mathrm{(eff)}}}\xspace}       
\def\Ope#1    {\ensuremath{\mathcal{O}_{#1}}\xspace}                       
\def\Opep#1   {\ensuremath{\mathcal{O}_{#1}^{'}}\xspace}                    

\newcommand{\ket}[1]{\ensuremath{|#1\rangle}}              

\newcommand{\aunit}[1]{\ensuremath{\text{\,#1}}}       

\newcommand{\tev}{\aunit{Te\kern -0.1em V}\xspace}
\newcommand{\gev}{\aunit{Ge\kern -0.1em V}\xspace}
\newcommand{\mev}{\aunit{Me\kern -0.1em V}\xspace}
\newcommand{\kev}{\aunit{ke\kern -0.1em V}\xspace}
\newcommand{\ev}{\aunit{e\kern -0.1em V}\xspace}
 
\newcommand{\mevc}{\ensuremath{\aunit{Me\kern -0.1em V\!/}c}\xspace}
\newcommand{\gevc}{\ensuremath{\aunit{Ge\kern -0.1em V\!/}c}\xspace}
\newcommand{\mevcc}{\ensuremath{\aunit{Me\kern -0.1em V\!/}c^2}\xspace}
\newcommand{\gevcc}{\ensuremath{\aunit{Ge\kern -0.1em V\!/}c^2}\xspace}

\def\gsim{{~\raise.15em\hbox{$>$}\kern-.85em
          \lower.35em\hbox{$\sim$}~}\xspace}
\def\lsim{{~\raise.15em\hbox{$<$}\kern-.85em
          \lower.35em\hbox{$\sim$}~}\xspace}

\def\tell1  {TELL1\xspace}
\def\ukl1   {UKL1\xspace}

\usepackage{cite} 
\usepackage{mciteplus}

\usepackage{longtable} 
\usepackage{caption}
\usepackage{subcaption}
\begin{document}

\renewcommand{\thefootnote}{\fnsymbol{footnote}}
\setcounter{footnote}{1}


\begin{titlepage}
\pagenumbering{roman}

\vspace*{-1.5cm}
\centerline{\large EUROPEAN ORGANIZATION FOR NUCLEAR RESEARCH (CERN)}
\vspace*{1.5cm}
\noindent
\begin{tabular*}{\linewidth}{lc@{\extracolsep{\fill}}r@{\extracolsep{0pt}}}
\ifthenelse{\boolean{pdflatex}}
{\vspace*{-1.5cm}\mbox{\!\!\!\includegraphics[width=.14\textwidth]{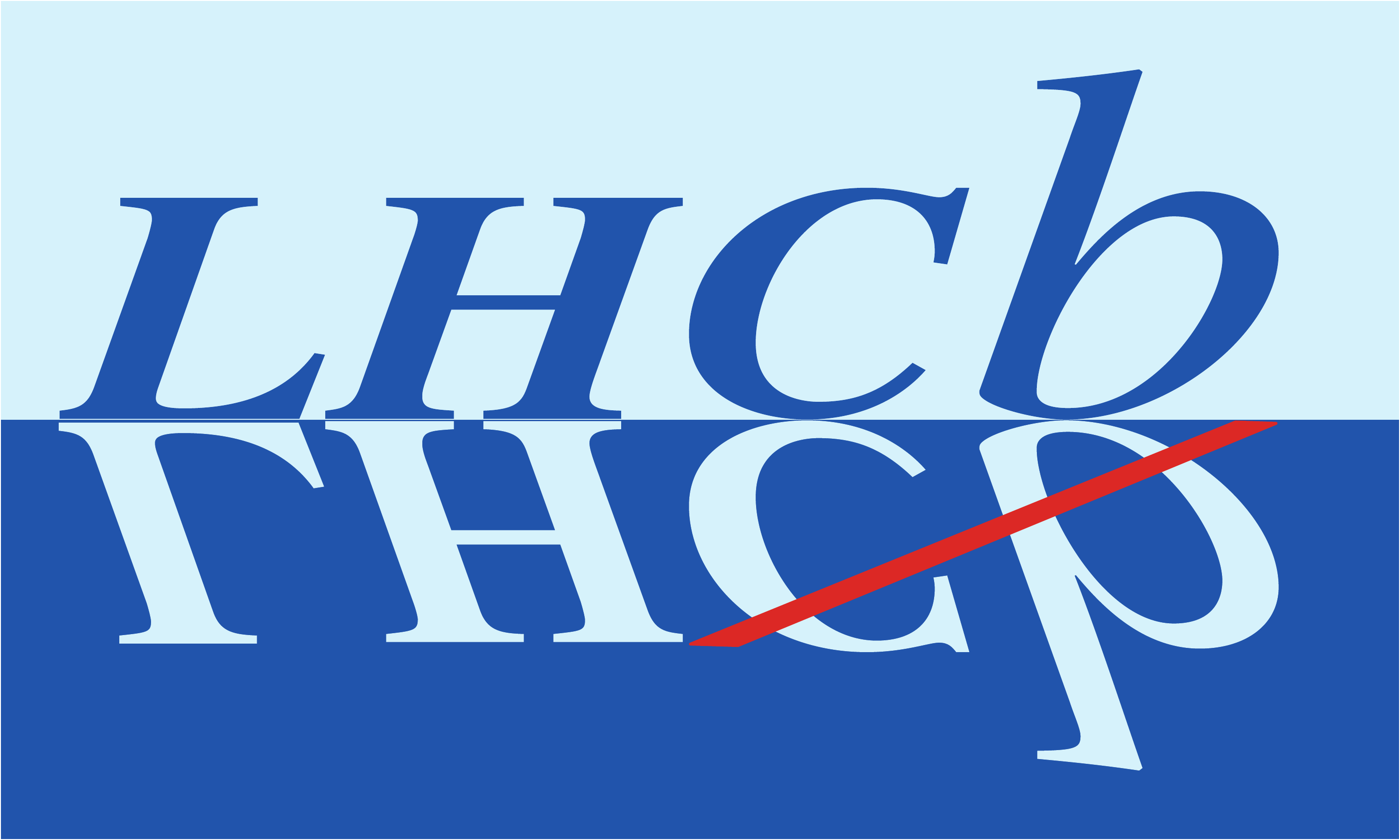}} & &}%
{\vspace*{-1.2cm}\mbox{\!\!\!\includegraphics[width=.12\textwidth]{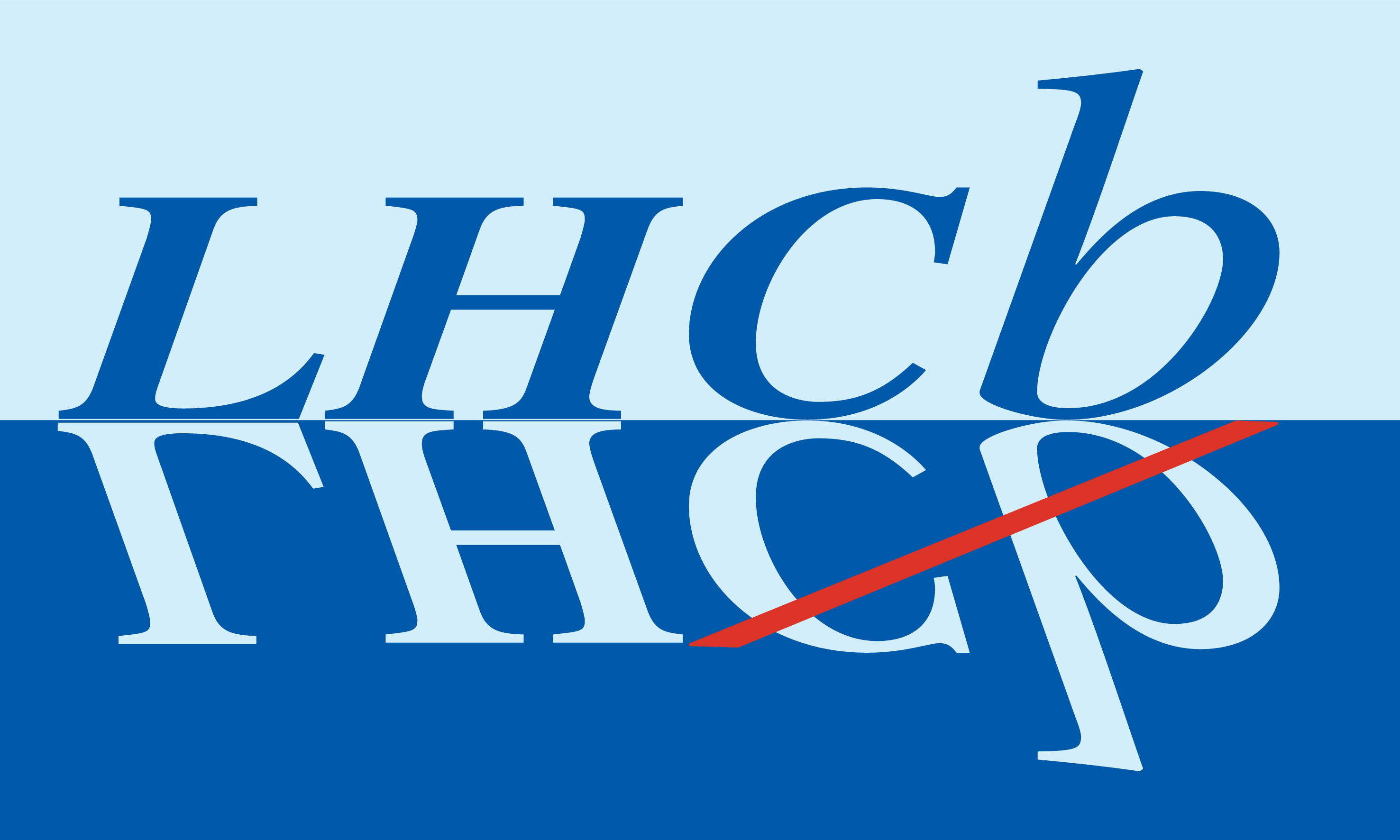}} & &}%
\\
 & &  \\  
 & & August 03, 2022 \\ 
 & & \\
\end{tabular*}

\vspace*{4.0cm}

{\normalfont\bfseries\boldmath\huge
\begin{center}
  \papertitle 
\end{center}
}

\vspace*{2.0cm}


\begin{center}
\paperauthors
\bigskip\\
{\normalfont\itshape\footnotesize
$ ^1$INFN Sezione di Padova, Padova, Italy\\
$ ^2$Nikhef National Institute for Subatomic Physics, Amsterdam, Netherlands \\
$ ^3$Università degli Studi di Padova, Padova, Italy\\
$ ^4$Universiteit Maastricht, Maastricht, Netherlands\\
$ ^5$Oliver Lodge Laboratory, University of Liverpool, Liverpool, United Kingdom \\
$ ^6$European Organization for Nuclear Research (CERN), Geneva, Switzerland \\
}
\end{center}

\vspace*{0.5cm}

\begin{abstract}
  \noindent
Machine Learning algorithms have played an important role in hadronic jet classification problems. The large variety of models applied to Large Hadron Collider data has demonstrated that there is still room for improvement. In this context Quantum Machine Learning is a new and almost unexplored methodology, where the intrinsic properties of quantum computation could be used to exploit particles correlations for improving the jet classification performance. In this paper, we present a brand new approach to identify if a jet contains a hadron formed by a $b$ or $\Bar{b}$ quark at the moment of production, based on a Variational Quantum Classifier applied to simulated data of the LHCb experiment. Quantum models are trained and evaluated using LHCb simulation. The jet identification performance is compared with a Deep Neural Network model to assess which method gives the better performance.  
  
\end{abstract}

\vspace*{0.5cm}

\begin{center}
  Published in
  JHEP 08 (2022) 014
\end{center}

\vspace{\fill}

{\footnotesize 
\centerline{\copyright~\papercopyright. \href{\paperlicenceurl}{\paperlicence}.}}
\vspace*{2mm}

\end{titlepage}


\newpage
\setcounter{page}{2}
\mbox{~}
%
%
%
%


\renewcommand{\thefootnote}{\arabic{footnote}}
\setcounter{footnote}{0}

\cleardoublepage


\pagestyle{plain} 
\setcounter{page}{1}
\pagenumbering{arabic}



\section{Introduction}
\label{sec:intro}

Machine Learning (ML) methods are widely used in experimental particle physics \cite{ref:hepml} data manipulation. One of the most successful applications is the classification of hadronic jets at the Large Hadron Collider (LHC) experiments \cite{ref:jetml}.
Jets are streams of particles produced via fragmentation and hadronization of quarks and gluons that emerge from particle collisions, \emph{e.g.} proton-proton collisions at LHC. They are complex objects formed by many detectable  particles, and it is possible to identify their properties by exploiting particle content and correlations, commonly referred to as jet substructure. Typical jet classification problems are the identification of the heavy-flavour hadron produced in the jet hadronization (\emph{e.g.} $b$ hadron vs $c$ hadron) or the identification of the charge of the heavy-flavour quark that constitutes this hadron (\emph{e.g.} $b$ vs $\bar{b}$).
State-of-the-art ML methods, such as Deep Neural Networks (DNN) \cite{ref:jetdnn}, Convolutional Neural Networks (CNNs)~\cite{ref:CNN}, Recurrent Neural Networks (RNNs)~\cite{ref:RNN}, Tensor Networks \cite{ref:ttnhep} and Graph Neural Networks \cite{ref:gnn,ref:pcloud}, have been applied to jets data collected by the LHC experiments, with a clear improvement of the classification performance with respect to classical non-ML methods~\cite{ref:deepjet,ref:atlasbtag,ref:atlasbcharge}.

Recently, Quantum Computing (QC) has set the scene for a revolution in ML. The new approach consists of using quantum circuits to tackle classification tasks, in the framework of Quantum Machine Learning (QML) \cite{ref:qml}: data are embedded into a quantum state, which is then passed to a variational quantum circuit, and by varying the circuit parameters a training procedure is performed by means of minimising a classical loss function. Probability measurements of the final state are then used to perform the classification.
Given the intrinsic properties of quantum computation, namely superposition and entanglement, the new approach could lead to new insights from the classification point of view. Jets that originate from gluons, or quarks of a certain charge and flavor, would have a characteristic particle content and correlations between them, which could be exploited to aid the identification of the original particle. It is interesting to study if QML, by exploiting the quantum nature of the algorithm, could enhance the classification performance.

QML techniques have recently been applied to solve High Energy Physics (HEP) problems, such as signal versus background separation \cite{ref:qmlappllhiggs,ref:qmlhiggs,ref:qmllhc1,ref:qmllhc2}, anomaly detection \cite{ref:qoptjets}, and particle track reconstruction \cite{ref:qmlappltracking,ref:track_luxe}. A more detailed review of QML applications to HEP can be found in Ref. \cite{ref:qmlreview}. This paper presents the first application of QML to the task of jet flavour identification. QML methods are performed on simulators and applied to simulated LHCb samples, to identify the charge of the $b$ quark that forms the $b$ hadron produced in the jet hadronization. In the rest of the paper, this task is simply referred as $b$-jet charge tagging.

The paper is structured in the following way: Sec.~\ref{sec:LHCb} provides a description of the LHCb jet reconstruction and identification together with the used dataset. In Sec.~\ref{sec:QuantumML} the considered QML algorithms are presented while the analysis flow is described in Sec.~\ref{sec:analysis}. The results are discussed in Sec.~\ref{sec:results}. The conclusions and future developments are presented in Sec.~\ref{sec:conclusions}.

\section{Jet reconstruction and identification at LHCb}
\label{sec:LHCb}

LHCb~\cite{ref:lhcb} is a single-arm spectrometer designed to study $b$ and $c$ hadrons in the forward region of proton-proton collisions. The reference system used at LHCb is defined by the $z$-axis (the beam axis) parallel to the proton beam, the $x$-axis parallel to the gravity acceleration, and the $y$-axis perpendicular to the other two. 
The direction of particle momentum is identified by the angles $\theta$ and $\phi$, where $\theta$ is the angle between the momentum and the $z$-axis, and $\phi$ is the azimuthal angle between the projection of the momentum in the $xy$ plane and the $y$-axis.
The pseudorapidity $\eta$ is defined as $\eta= -\mathrm{log} \left [ \mathrm{tan} \left (\frac{\theta}{2} \right)\right ]$.
The LHCb detector covers the region in the pseudorapidity range $2 < \eta < 5$, and consists of a tracking system and a particle identification system~\cite{ref:lhcbperformance}.
The tracking system is formed by a vertex detector, several tracking stations and a dipole magnet. The vertex detector efficiently reconstructs the decay vertex of $b$- and $c$-hadron decays, while the tracking stations measure the trajectories (tracks) of charged particles and their momenta.
The particle identification system is formed by two Ring Imaging Cherenkov detectors, two calorimeters (electromagnetic and hadronic) and a muon detector, that allows to precisely determine the type of the particles produced in the collision. Jets are reconstructed using inputs selected by the Particle Flow algorithm~\cite{ref:partflow}. These are charged particles detected by the tracking system, and neutral particles reconstructed in the calorimeter system as energy clusters isolated from tracks~\cite{ref:lhcbjets}. 
The selected particles represent the input of the anti-$k_t$ clustering algorithm~\cite{ref:antikt} for jet clusterization. The jet clusterization is done using \textsc{FastJet}~\cite{ref:fastjet} with radius $R=0.5$. The jet momentum is defined as the sum of the momenta of the particles forming the jet. The jet axis is defined as the direction of the jet momentum. 
Particles inside a jet are approximately contained in a cone structure with a distance from the jet axis $\Delta R=\sqrt{(\Delta \eta)^2+(\Delta \phi)^2}=0.5$, where $\Delta \eta$ is the difference in pseudo-rapidity and $\Delta \phi$ is the difference in the azimuthal angle with respect to the jet axis. 

The goal is to distinguish between jets that contain a $b$ or $\bar{b}$ hadron just after the hadronization, \emph{i.e.} in the instant of $b$-hadron production and not at decay, since neutral $B$-mesons can undergo flavour oscillation. Therefore the analysis is restricted to a sample of jets that belong to these two categories, labelled as $b$ jets and $\bar{b}$ jets.

This preliminary selection is performed in two steps:
\begin{itemize}
    \item reconstruction of a vertex (secondary vertex) significantly displaced from the primary proton-proton interaction point, using tracks detected by the vertex detector~\cite{ref:svtag}, representing the $b$-hadron decay point;
    \item identification of the jet that contains the secondary vertex within its cone.
\end{itemize}
The $b$-jet charge tagging subsequently becomes a binary classification problem where the jet can belong to one of the two exclusive categories: $b$ jets or a $\bar{b}$ jets. 
The charge of the $b$ quark at production is correlated to the charge of the $b$-hadron decay products. 
This correlation is not perfect, since neutral $B$-mesons can oscillate, and the charge of the $b$ quark at production may be different from the charge at decay.
As an example, in semi-leptonic decays, the $b$ hadron can produce a muon, whose charge is directly related to the charge of the $b$ quark. 
However, due to possible $B$-meson oscillations and to the large number of particles, including muons, in a jet the information is diluted and that has to be taken into account. 

\begin{figure}[tb]
\centering
\includegraphics[width=3.2in]{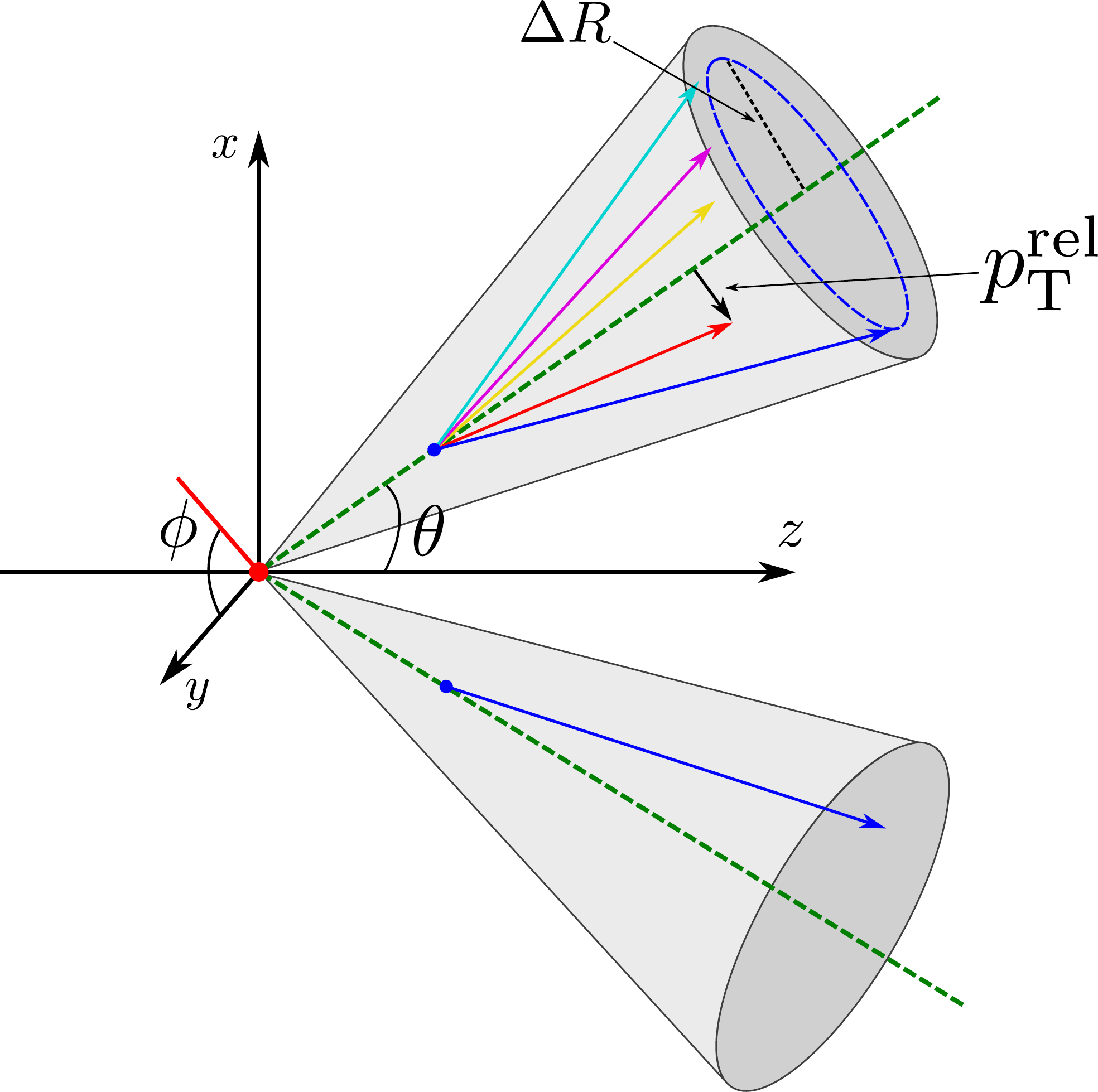}
\caption{Sketch representing possible jet tagging methods. In the exclusive method the information comes from a particle, e.g. the muon, whose charge is correlated to the $b$ hadron  (lower jet); in the inclusive method, the information is extracted from the jet constituents (upper jet). The magnitude of the particle momentum transverse to the jet axis is labelled as $p^{\text{rel}}_{\mathrm{T}}$.}
\label{fig:Btagging}
\end{figure}

Two types of $b$-jet charge tagging algorithms are currently used in LHCb:
\begin{itemize}
    \item \emph{exclusive algorithms}, based on information coming from particles inside the jet strictly correlated with the $b$-hadron decay, such as the muon;
    \item and \emph{inclusive algorithms}, which aim to exploit  the jet sub-structure, \textit{i.e.} information coming from the jet constituents, as shown in Fig.~\ref{fig:Btagging}.
    \end{itemize}
The QML approach presented in this paper belongs to the category of \emph{inclusive algorithms}.
However, for the sake of comparison, the results are compared to the \emph{muon tagging}~\cite{ref:asymmetry}, which is an exclusive algorithm.  This tagger selects the muon with the highest momentum with respect to the $z$ axis, $p_{\mathrm{T}}$, inside the jet. This simple requirement is sufficient to identify the muon coming from a $b$ hadron and therefore to infer the quark charge by measuring the muon charge. The efficiency of this algorithm is limited by the probability that  a $b$ hadron decays to a final state with a muon, which is $\sim 10 \%$~\cite{ref:pdg}.

The performance of different $b$-jet charge tagging algorithms are compared using the tagging power $\epsilon_{\text{tag}}$~\cite{tag_pow1,tag_pow2,tag_pow3}, defined as
\begin{equation}
\label{eq:tagging_power}
    \epsilon_{\text{tag}}=\epsilon_{\text{eff}}(2a-1)^2
\end{equation}
as the figure of merit, where $\epsilon_{\text{eff}}$ is the tagging efficiency, \emph{i.e.} the fraction of jets where the classifier takes a decision, and $a$ is the accuracy, \emph{i.e.} the fraction of correctly tagged jets with respect to the tagged jets.
The tagging power is the effective fraction of events that contribute to the statistical uncertainty in a measurement where the $b$-jet charge tagging is applied.

\subsection{Data samples description}
\label{subsec:LHCbdataset}
LHCb simulated samples are used in the studies presented in this paper. The $b\bar{b}$ di-jets samples are produced within the LHCb simulation framework~\cite{ref:gauss}, which  uses \textsc{Pythia8.1}~\cite{ref:pythia} with a specific LHCb configuration~\cite{Belyaev_2011}, to generate proton-proton interactions and jet fragmentation and hadronization at a center-of-mass energy of $13\,$TeV, and an internal implementation of \textsc{EvtGen}~\cite{ref:evtgen} to simulate $b$ hadron decays. The \textsc{Geant4} software~\cite{ref:geant4}, embedded in the LHCb framework, is used to simulate the detector response. 
Pairs of  $b$ and $\bar{b}$ jets are selected by requiring for each jet a  $p_{\mathrm{T}}$ greater than 20 GeV/c and a $\eta$ in the range $2.2<\eta<4.2$, to ensure that they are well inside the instrumented part of the detector.
After the pre-selection, a fixed number of $16$ different features related to the jet substructure are used as input to the classifiers. Among the reconstructed particles inside a jet the muon, kaon, pion, electron and proton with the highest $p_\mathrm{T}$ are selected.  For each particle three physical variables are considered: the magnitude of the transverse momentum to the jet axis ($p^{\text{rel}}_{\mathrm{T}}$), the charge ($q$), and the distance, measured in the ($\eta$,$\phi$) space, between the particle and the jet axis ($\Delta R$). If a particle type is missing, the relative features are set to $0$.
The last feature is the weighted jet charge $Q$, defined as the sum of the charges of the particles inside the jet weighted with the particles $p^{\text{rel}}_{\mathrm{T}}$~\cite{ref:jetcharge1,ref:jetcharge2,ref:atlasjetchargestudies,ref:atlasjetvertexcharge}
\begin{equation}
Q =  \frac{\sum_i (p^{\text{rel}}_{\mathrm{T}})_i q_i}{\sum_i (p^{\text{rel}}_{\mathrm{T}})_i} ~ .
\end{equation}

The analysis for the $b$-jet identification is performed with two datasets. The \textit{complete dataset} includes the events selected with the 16 features described above. The \textit{muon dataset} contains jets with at least one muon and only four features: $p_{\mathrm{T}}^{\text{rel}}$, $\Delta R$, $q$ of the muon and the weighted jet charge $Q$. Table~\ref{tab:datasets} summarises the characteristics of the data samples.
\begin{table}[]
\resizebox{\textwidth}{!}{
\begin{tabular}{c|ccc|ccc|ccc|ccc|ccc|p{0.6cm}|}
\cline{2-17}
 & \multicolumn{3}{c|}{\textbf{Muon}} & \multicolumn{3}{c|}{\textbf{Kaon}} & \multicolumn{3}{c|}{\textbf{Pion}} & \multicolumn{3}{c|}{\textbf{Electron}} & \multicolumn{3}{c|}{\textbf{Proton}} & \\ \cline{2-17} 
Dataset & $p_\mathrm{T}^\mathrm{rel}$ & $q$ & $\Delta R$ & $p_\mathrm{T}^\mathrm{rel}$ & $q$ & $\Delta R$ & $p_\mathrm{T}^\mathrm{rel}$ & $q$ & $\Delta R$ & $p_\mathrm{T}^\mathrm{rel}$ & $q$ & $\Delta R$ & $p_\mathrm{T}^\mathrm{rel}$ & $q$ & $\Delta R$ &\hspace{0.05cm} $Q$ \\ \hline
\multicolumn{1}{|c|}{Complete} & \multicolumn{1}{p{0.6cm}|}{\checkmark} & \multicolumn{1}{p{0.6cm}|}{\checkmark} & \checkmark & \multicolumn{1}{p{0.6cm}|}{\checkmark} & \multicolumn{1}{p{0.6cm}|}{\checkmark} & \checkmark & \multicolumn{1}{p{0.6cm}|}{\checkmark} & \multicolumn{1}{p{0.6cm}|}{\checkmark} & \checkmark & \multicolumn{1}{p{0.6cm}|}{\checkmark} & \multicolumn{1}{p{0.6cm}|}{\checkmark} & \checkmark & \multicolumn{1}{p{0.6cm}|}{\checkmark} & \multicolumn{1}{p{0.6cm}|}{\checkmark} & \checkmark & \checkmark \\ \hline
\multicolumn{1}{|c|}{Muon} & \multicolumn{1}{p{0.6cm}|}{\checkmark} & \multicolumn{1}{p{0.6cm}|}{\checkmark} & \checkmark & \multicolumn{1}{p{0.6cm}|}{} & \multicolumn{1}{p{0.6cm}|}{} &  & \multicolumn{1}{p{0.6cm}|}{} & \multicolumn{1}{p{0.6cm}|}{} &  & \multicolumn{1}{p{0.6cm}|}{} & \multicolumn{1}{p{0.6cm}|}{} &  & \multicolumn{1}{p{0.6cm}|}{} & \multicolumn{1}{p{0.6cm}|}{} &  & \checkmark \\ \hline
\end{tabular}%
}
\caption{Summary of the features contained in the two datasets.}
\label{tab:datasets}
\end{table}

\section{Quantum Machine Learning models}
\label{sec:QuantumML}
A quantum algorithm is implemented by means of a quantum circuit, namely a collection of linked quantum gates acting on a $n$-qubit quantum state: the measurements on the final state represent the outcome of the quantum algorithm. Parametrized Quantum Circuits (PQCs)~\cite{ref:pqc} are a type of circuit that contains adjustable gates with tunable parameters. The Variational Quantum Classifier (VQC)~\cite{ref:vqc} is a hybrid quantum-classical algorithm to perform classification tasks using a Machine Learning model based on a PQC with the following structure:
\begin{description}
    \item [Data encoding] data $x$, the features representing the jet substructure in this application, are pre-processed and encoded into a subset of the parameters of a PQC. The stage produces a quantum state $\ket{x}$ representing the input jet.
    \item[Variational circuit] the state $\ket{x}$ is processed by a PQC, $U(\theta)$, featuring trainable parameters $\theta$ to be optimised during the training phase. This stage produces a final state $\ket{\psi} = U(\theta)\ket{x}$.
    \item[Prediction] expectation values computed on the final state $\ket{\psi}$ are mapped to probabilities for the two labels, $P_b$ and $P_{\bar{b}}$. The training process aims to match the label predictions with the true charge of the $b$-hadron in the jet in the instant of production, available in the simulations.
\end{description}
Two different PQC models are studied in this work: Amplitude Embedding and Angle Embedding, described below.
\label{sec:qmlmodels}
\subsection{Amplitude Embedding}
\label{subsec:amplitudeembedding}
\begin{figure}
    \centering
    \includegraphics[width=0.8\textwidth]{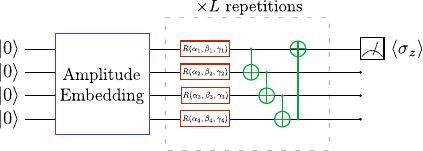}
    \caption{Circuit representation of the Amplitude Embedding model. In blue, variables are embedded into the amplitudes of a quantum state. In red, trainable generic rotational gates to be optimised during the training phase. In green, CNOT gates entangling qubits with a circular topology.}
    \label{fig:amplitude_embedding_circuit}
\end{figure}

The Amplitude Embedding model consists in a PQC made by an embedding circuit followed by a variational circuit. The schematic representation of this model is shown in Fig.~\ref{fig:amplitude_embedding_circuit}. The embedding circuit consists of an \textit{Amplitude Encoder} that encodes up to $2^n$ features into the amplitude of a $n$-qubit quantum state, or equivalently, a vector of $N$ features can be encoded using $\lceil\log_2 N\rceil$ qubits:
\begin{equation}
    \ket{x} = \sum_{i=1}^{2^n} x_i\,\ket{n_i}
\end{equation}
where $x_i$ is the $i^\text{th}$ feature and $\ket{n_i}$ is the $i^\text{th}$ vector of the computational basis. The definition requires the $x$ vector to be normalised,
\begin{equation}
    \sum_i |x_i|^2 = 1.
\end{equation}
If the number of features to encode is not a power of 2, the remaining amplitudes can be padded with constant values. This model embeds the 16 (4) variables of the \emph{complete dataset} (\emph{muon dataset}) into the amplitudes of a 4-qubit (2-qubit) quantum state. The variational stage consists of a variable number $L$ of \textit{strongly entangling layers}. A strongly entangling layer consists of trainable generic rotational gates $R(\alpha_i, \beta_i,\gamma_i)$ applied to each qubit followed by a collection of CNOT gates applied to neighbouring pairs of qubits, considering the last one as a neighbour of the first one. The complexity of this kind of circuit can be tuned by changing the number of strongly entangling layers $L$: for a generic $n$-qubit circuit, the number of trainable parameters of the model $N_\text{par}$ is equal to
\begin{equation}
\label{eq:se_npar}
    N_\text{par} = 3 \times n \times L.
\end{equation}
On the final state, the expectation of the Pauli operator of the first qubit $\langle \sigma^0_z \rangle \in [-1,+1]$ is measured and used to define the probabilities $P_b$ and $P_{\Bar{b}}$ of being a $b$-jet and a $\bar{b}$-jet, respectively:
\begin{align}
\label{eq:Pb}
    P_b &= \frac{1}{2}(\langle \sigma^0_z \rangle + 1) \\
\label{eq:Pbbar}
    P_{\Bar{b}} &= \frac{1}{2}(1-\langle \sigma^0_z \rangle) = 1-P_b.
\end{align}

\subsection{Angle Embedding}
\label{subsec:angleembedding}
\begin{figure}[tb]
    \centering
    \includegraphics[width=0.85\textwidth]{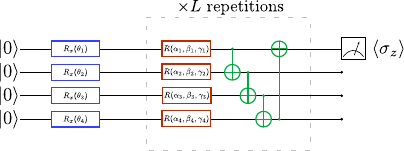}
    \caption{Circuit representation of the Angle Embedding model. In blue, $x$-axis rotational gates used to embed the variables into the quantum circuit. In red, trainable generic rotational gates to be optimised during the training phase. In green, CNOT gates entangling qubits with a circular topology.}
    \label{fig:angle_embedding_circuit}
\end{figure}

 The structure of the Angle Embedding model, represented in Fig.~\ref{fig:angle_embedding_circuit}, differs from the Amplitude Embedding model in the encoding used to embed the features of the datasets into the quantum state: in this case, the embedding circuit consists in a \textit{Angle Encoder} that embeds 16 (4) features of the \emph{complete dataset} (\emph{muon dataset}) as rotation angles of  16 (4) $x$-axis rotational gates $R_x(\theta_i)$. Therefore, this circuit structure requires a one-to-one correspondence between qubits and input features: that makes it impractical to adopt with high-dimensionality datasets, due to computational constraints of quantum simulators.
 The variational stage of the circuit is identical to the Amplitude Embedding model, featuring a variable number $L$ of strongly entangling layers that can be opportunely chosen to tune the number of parameters $N_\text{par}$, defined in Eq.~\ref{eq:se_npar}, and, therefore, the complexity. The measurement of the expectation value of the Pauli $\sigma_z$ operator is mapped to the tagging probabilities $P_b$ and $P_{\bar{b}}$ as expressed in Eq.~\ref{eq:Pb} and Eq.~\ref{eq:Pbbar}, identically to the Amplitude Embedding model.

\section{$\mathbf{b}$-jets identification  procedure}
\label{sec:analysis}
Quantum circuits are simulated by means of noiseless simulators (noise impact is studied in Sec.~\ref{subsec:noisemodels}) using Pennylane~\cite{ref:pennylane}, a Python framework designed specifically for QML applications. The quantum circuit is embedded into a classical optimisation algorithm, using the Jax~\cite{ref:jax} Python library. Since the quantum algorithms results are compared to classical DNN ones,  the same analysis is performed with a standard feed-forward deep neural network, implemented using the Keras~\cite{ref:keras} framework with the TensorFlow~\cite{ref:tensorflow} back-end. Additional details on the structure and the optimisation of the DNN are reported in App.~\ref{app:DNN}.

\subsection{Training and testing phases}
\label{subsec:traintest}
\begin{figure}[tb]
    \centering
    \begin{subfigure}[b]{0.49\textwidth}
    \centering
    \includegraphics[width=\textwidth]{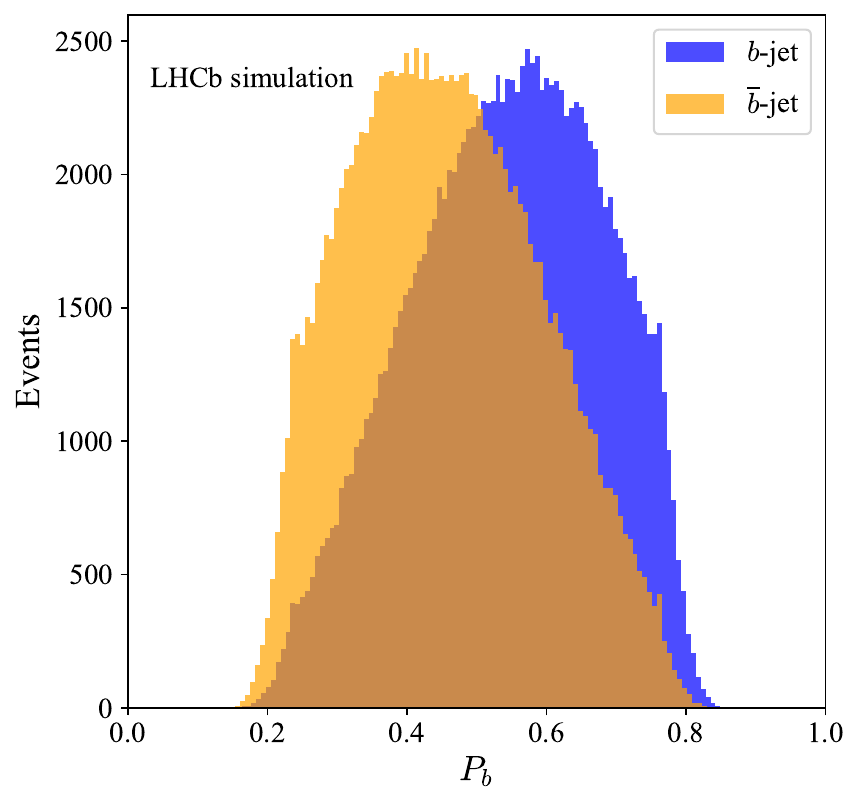}
    \caption{}
    \label{fig:prob_dist_left}
    \end{subfigure}
    \hfill
    \begin{subfigure}[b]{0.49\textwidth}
    \centering
    \includegraphics[width=\textwidth]{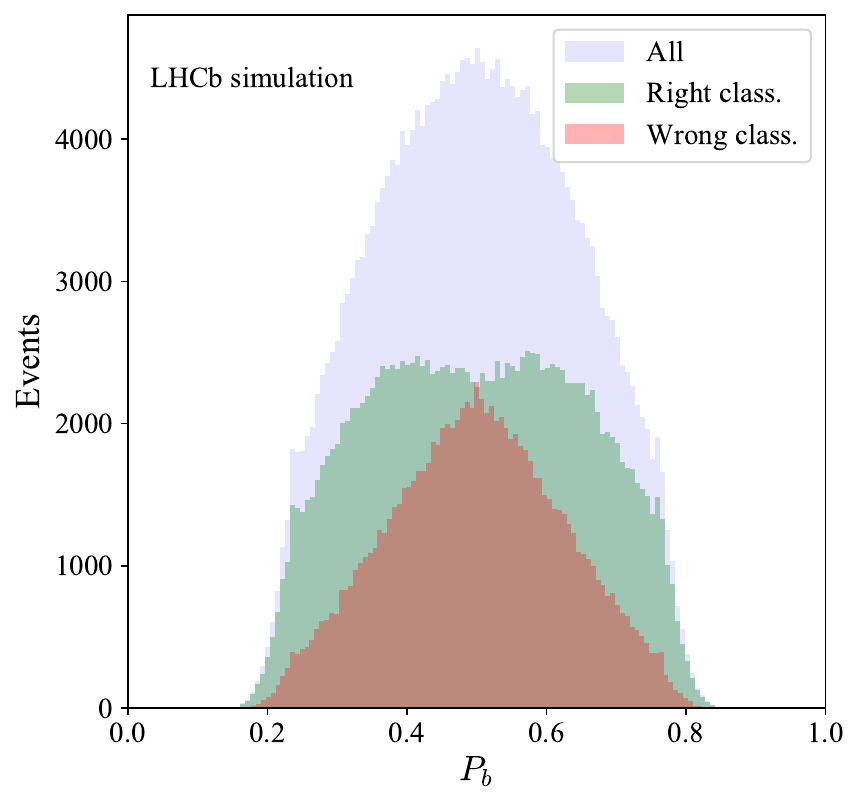}
    \caption{}
    \label{fig:prob_dist_right}
    \end{subfigure}
    
    \caption{Probability distributions for jet tagged to $b$ (blue) and $\bar{b}$ quarks (yellow), showing separation around 0.5 (a). Probability distribution for the Angle Embedding circuit: jet correctly (wrongly) tagged are plotted in green (red), showing around 0.5 worse classification. The probability distribution for all jets is shown in grey (b).}
    \label{fig:prob_dist}
\end{figure}

\begin{figure}[tb]
    \centering
    \begin{subfigure}[b]{0.49\textwidth}
    \centering
    \includegraphics[width=\textwidth]{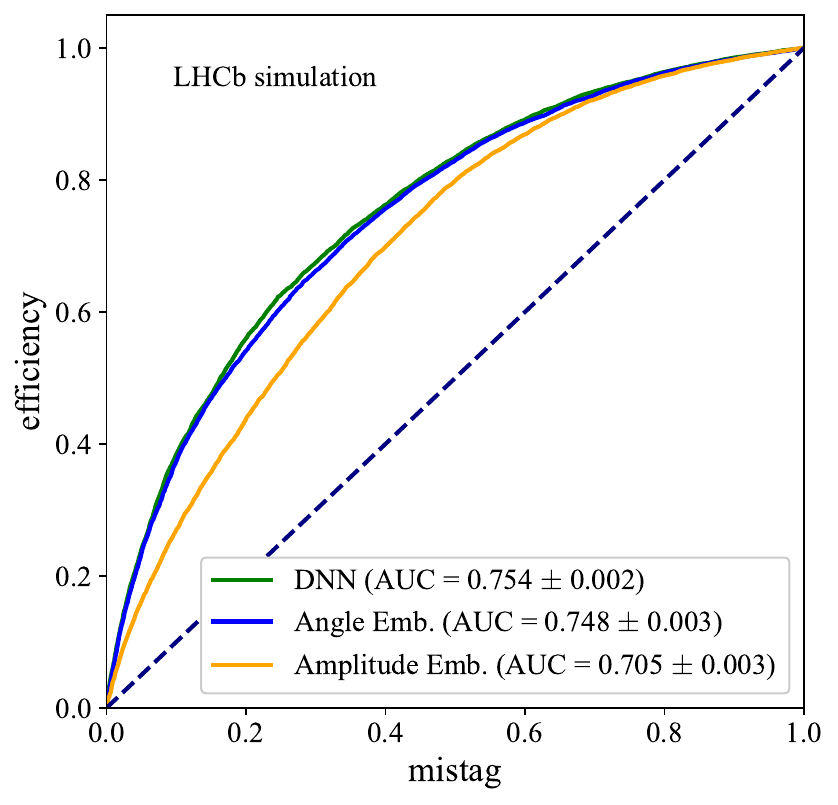}
    \caption{}
    \label{fig:ROC_mu}
    \end{subfigure}
    \hfill
    \begin{subfigure}[b]{0.49\textwidth}
    \centering
    \includegraphics[width=\textwidth]{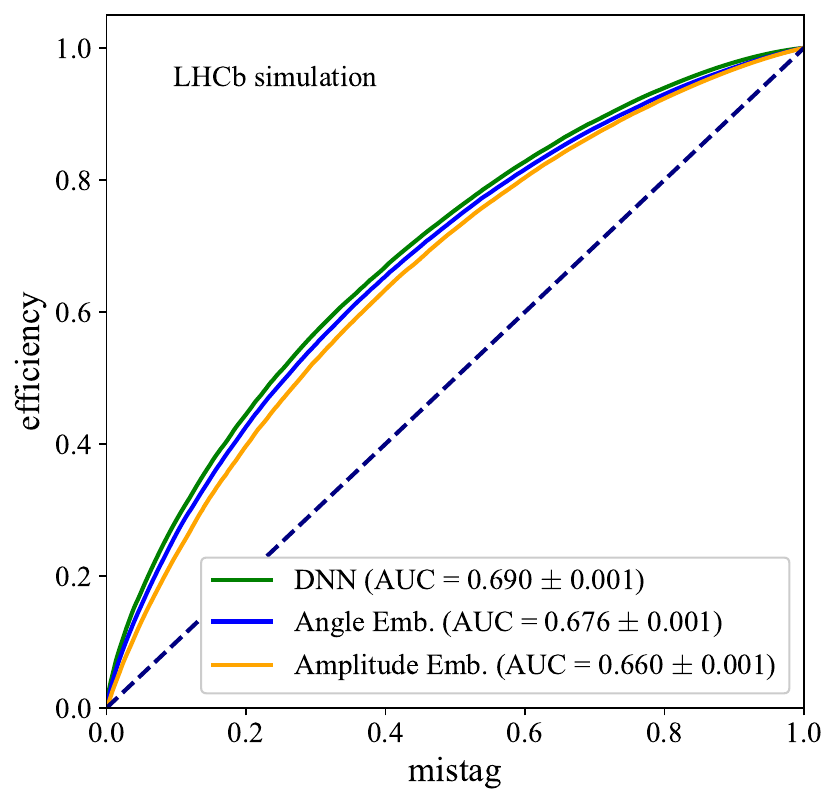}
    \caption{}
    \label{fig:ROC_complete}
    \end{subfigure}
    \caption{ROC distributions and AUC score for DNN (green), Angle Embedding (blue) and Amplitude Embedding circuits (yellow) for the \textit{muon dataset} (a) and the \textit{complete dataset} (b). The dashed line represents a random classifier.}
    \label{fig:ROC}
\end{figure}

The \emph{muon} and \emph{complete} datasets are both  split into training and testing sub-datasets: about $60\%$ of the samples are used in the training  process that includes also the validation and the remaining $40\%$ are used to test,  evaluate and compare the classifiers. 
In the \emph{muon dataset} analysis, 60000 jets are used for training and 40000 jets are used for testing.  The \emph{complete dataset} training is performed on 400000 jets and remaining 290000 are used for testing and assessing performance.
In the analysis of the \emph{muon dataset} (\emph{complete dataset}), the Angle Embedding and Amplitude Embedding classifiers are studied and compared to a DNN with the same 4 (16) input variables.
The training process aims to find the values of the model parameters $\theta$ that minimise the Mean Squared Error loss function
\begin{equation}
\label{eqn:lossfunction}
    \mathcal{L}(\theta) = \frac{1}{N}\sum_{i=1}^{N}(P^i_b(\theta)-T^i)^2,
\end{equation}
where $N$ is the number of training jets, $P^i_b$ is the predicted probability, defined in Eq.~\ref{eq:Pb}, for the $i$-th jet, and $T^i$ is the target probability for the $i$-th jet, \textit{i.e.}, $1$ for a $b$ jet and $0$ for a $\bar{b}$ jet.
Due to the large number of jets in the datasets, the quantum models are trained implementing a mini-batch gradient descent~\cite{BatchGD} algorithm using the ADAM optimiser~\cite{ref:adam} to minimise Eq.~\ref{eqn:lossfunction}. The training dataset is split in several mini-batches containing a fixed number of training samples. During each training step, the gradient of Eq.~\ref{eqn:lossfunction} is evaluated, averaging over the training samples of a mini-batch, and used to update model parameters. A training epoch is completed when the whole training dataset is processed, namely after a number of steps equal to the number of mini-batches. Unless specified otherwise, the models are trained with learning rate\footnote{In Machine Learning, the learning rate $\xi$ of an optimisation algorithm is usually defined as the scaling factor applied to the gradient of the loss function when updating the parameters. It can be tuned as a parameter of the learning process.} $\xi=0.01$ for 100 epochs, while the mini-batch size is fixed to the maximum value allowed by memory constraints. 
The output of the classifier gives the probability that a jet is generated by a $b$ or a $\Bar{b}$ quark. The label with the highest probability is assigned to the jet, \textit{i.e.} if $P_b > 0.5$ ($P_b < 0.5$) then it is classified as a $b$ jet ($\Bar{b}$ jet).
In Fig.~\ref{fig:prob_dist_left} the output distributions for the two classes ($b$ and $\bar{b}$ jets) after the training procedure are shown; a separation between the two distributions around $0.5$ is visible  leading to a good classification. It should be noted that the $P_b$ distribution is shifted toward 1 for $b$ quarks, and toward 0 for $\bar{b}$ quarks, as expected.
Fig.~\ref{fig:prob_dist_right} shows the output distribution for the Angle Embedding classifier on 16 qubits: in green the jets that are correctly classified,  in red the jets that are wrongly classified and the sum of all jets in grey. As expected, correctly classified jets tend to stay close to $0$ and $1$ while the wrongly classified jets are peaked around $0.5$, where the prediction power is minimum. 
Figure~\ref{fig:ROC} shows the Receiver Operating Characteristic curve (ROC) and the Area Under Curve (AUC) for the DNN and the quantum classifiers for the \textit{muon dataset} and the \textit{complete dataset}.

\section{Results on $b$-jet charge tagging}
\label{sec:results}
\begin{figure}[tb]
    \centering
    \includegraphics[width=0.65\textwidth]{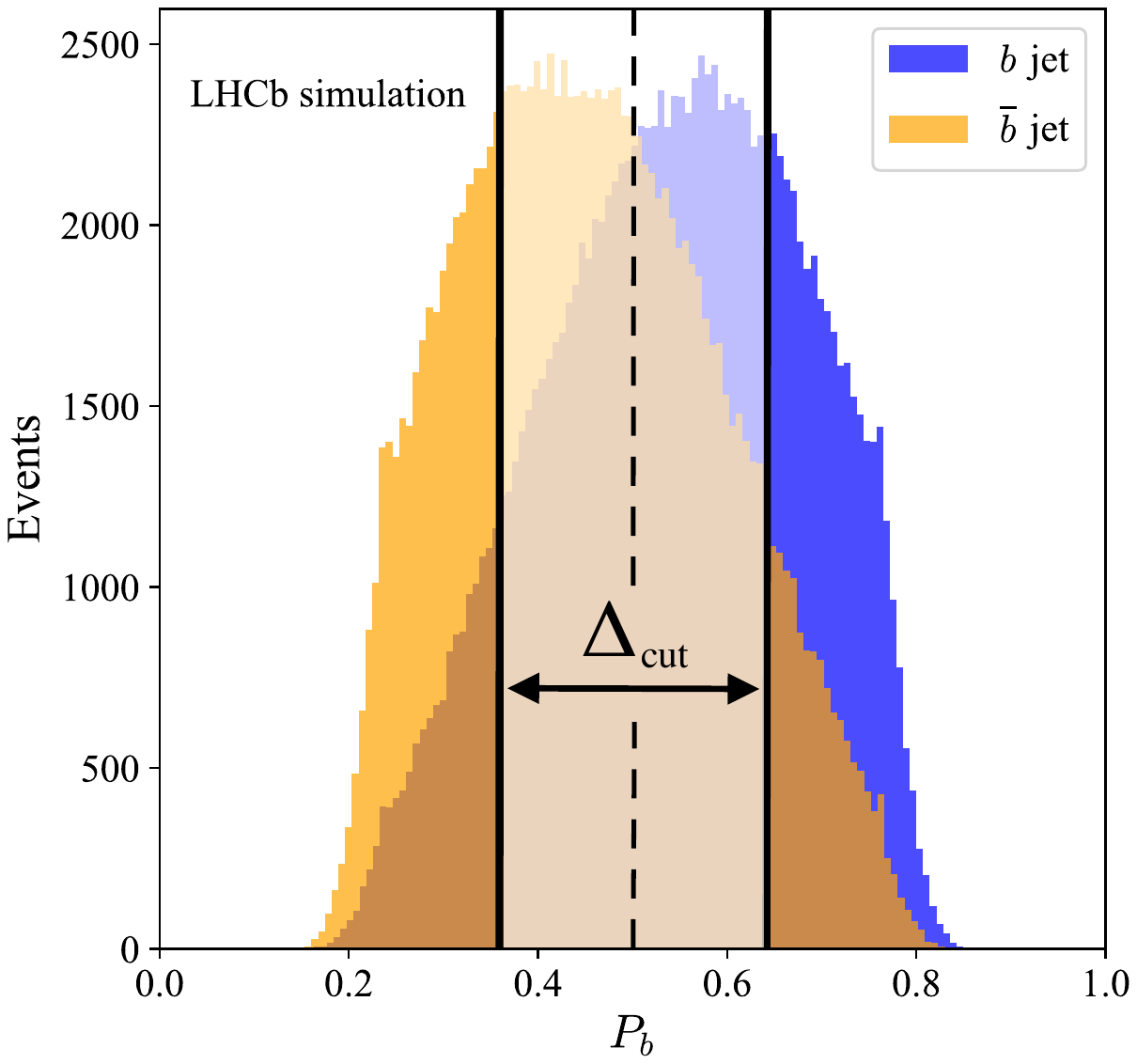}
    \caption{Probability distributions for jet tagged to (blue) $b$   and (yellow) $\bar{b}$ quarks }
    \label{fig:prob_dist_cut}
\end{figure}

The performance of the classifiers is evaluated  by using the jet tagging power, defined in Eq.~\ref{eq:tagging_power}. The tagging power is computed as a function of the jet $p_{\mathrm{T}}$ and $\eta$ for both the quantum and the classical classifiers. In order to optimise the tagging power, a region symmetric with respect to 0.5 is defined, where no classification is performed. The width $\Delta_{\text{cut}}$ of the excluded region is defined for each classifier by maximising the tagging power evaluated using all the jets in the dataset. Such an exclusion region reduces the tagging efficiency because less jets are tagged, but enhances the identification probability.
The probability distributions and the excluded region are shown in Fig.~\ref{fig:prob_dist_cut}: indeed the region where the prediction power is minimum is excluded. The width $\Delta_{\text{cut}}$ of the excluded region for each classifier and for \emph{muon} and \emph{complete dataset} are summarised in Tab.~\ref{table:delta_cut}. For comparison, the unoptimised tagging power, obtained by identifying as $b$ ($\bar{b}$) the jets with $P_b$ above (below) 0.5, is presented in App.~\ref{app:noOptTagPower}.

\begin{table}[tb]
\caption{Width $\Delta_{\text{cut}}$ for different classifiers and dataset.}
\label{table:delta_cut}
\centering
\begin{tabular}{c|ccc|}
\cline{2-4}
 & \multicolumn{3}{c|}{\textbf{Classifier}} \\
\cline{2-4}
\multicolumn{1}{c|}{Dataset} & DNN & Angle Embedding & Amplitude Embedding\\
\hline
\multicolumn{1}{|c|}{Muon} & 0.30 & 0.25 & 0.16\\
\hline
\multicolumn{1}{|c|}{Complete} & 0.21 & 0.19 & 0.12\\
\hline
\end{tabular}
\end{table}

\begin{figure}[tb]
    \centering
    \begin{subfigure}[b]{0.49\textwidth}
    \centering
    \includegraphics[width=\textwidth]{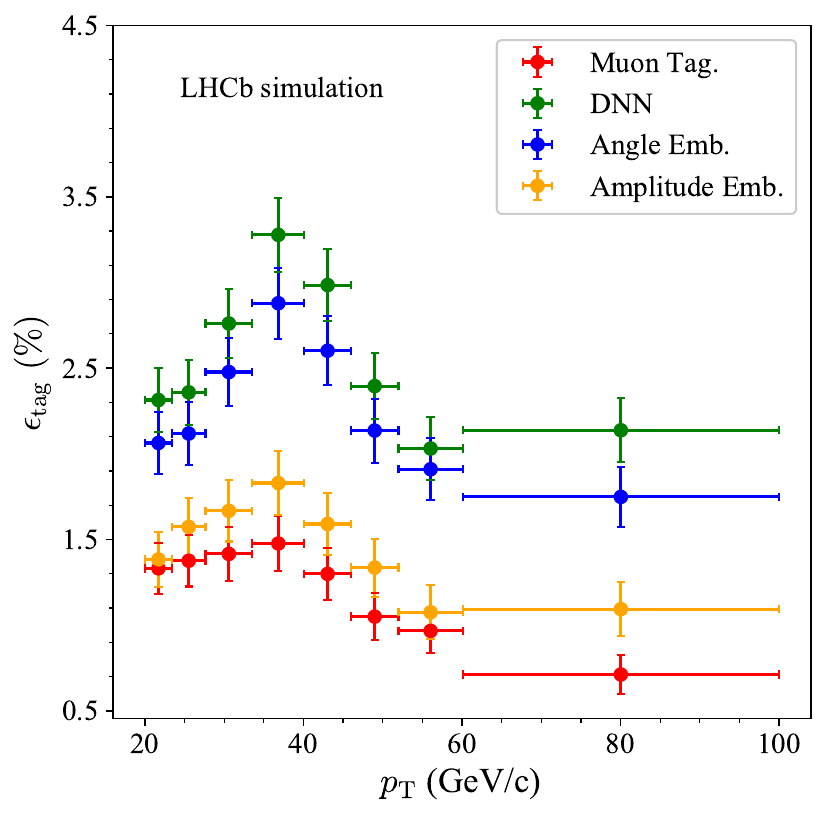}
    \caption{}
    \label{fig:tagpower_mu_left}
    \end{subfigure}
    \hfill
    \begin{subfigure}[b]{0.49\textwidth}
    \centering
    \includegraphics[width=\textwidth]{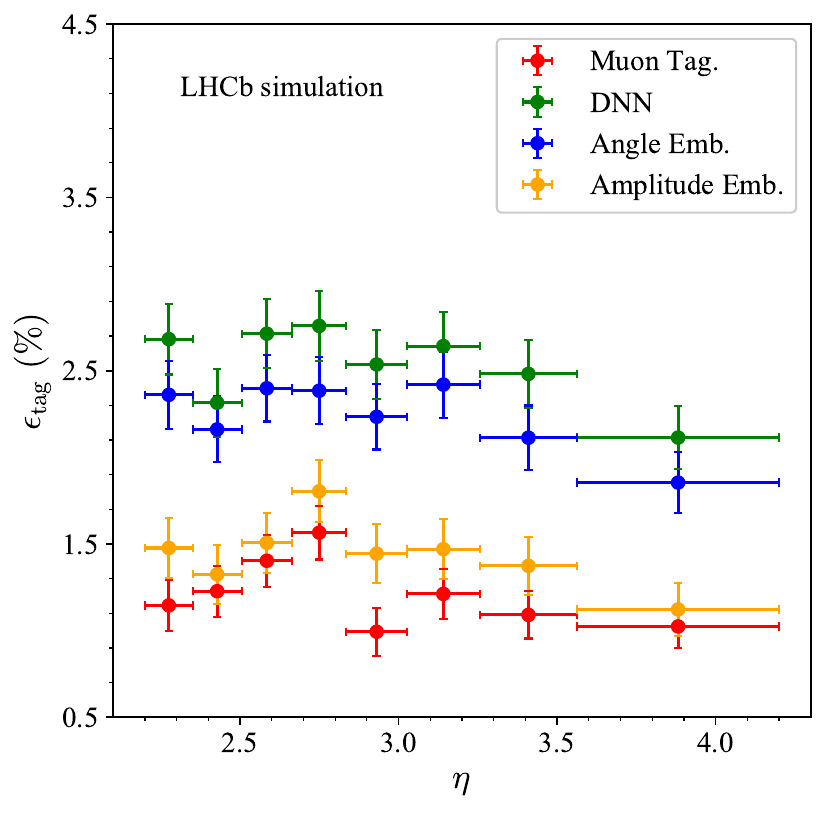}
    \caption{}
    \label{fig:tagpower_mu_right}
    \end{subfigure}
     
    \caption{Tagging power $\epsilon_{\text{tag}}$ with respect to (a) jet $p_{\mathrm{T}}$  and (b) jet $\eta$  for the \emph{muon dataset}. The Angle Embedding circuit and the DNN show similar performance.}
    \label{fig:tagpower_mu}
\end{figure}

The tagging power $\epsilon_{\text{tag}}$ as a function of jet $p_{\mathrm{T}}$ and $\eta$ for the classical and quantum classifiers applied to the \emph{muon dataset} is shown in Fig.~\ref{fig:tagpower_mu}. All the distributions have similar behaviour demonstrating that no bias is created by any algorithm. The tagging power dependence on the jet $p_{\mathrm{T}}$ is as expected, since at high $p_{\mathrm{T}}$ the reconstruction and identification of the jet particle content is more difficult, leading to a lower tagging power ~\cite{ref:btag}. The DNN shows slightly better performance compared to the Angle Embedding algorithm, even though the two results are compatible within $2\,\sigma$ in each $p_\mathrm{T}$ and $\eta$ bin. Both algorithms outperform the Amplitude Embedding model and the muon tagging approach. The muon tagging performance is expected, since it only uses the muon charge $q$ for the prediction. The other algorithms use also the muon $p_{\mathrm{T}}$, the muon $\Delta R$ and the weighted jet charge $Q$. The simple 2-qubit Amplitude Embedding model shows a slightly better performance with respect to the muon tagging, but still worse than the DNN and the Angle Embedding model. 

\begin{figure}[ht]
    \centering
    \begin{subfigure}[b]{0.49\textwidth}
    \centering
    \includegraphics[width=\textwidth]{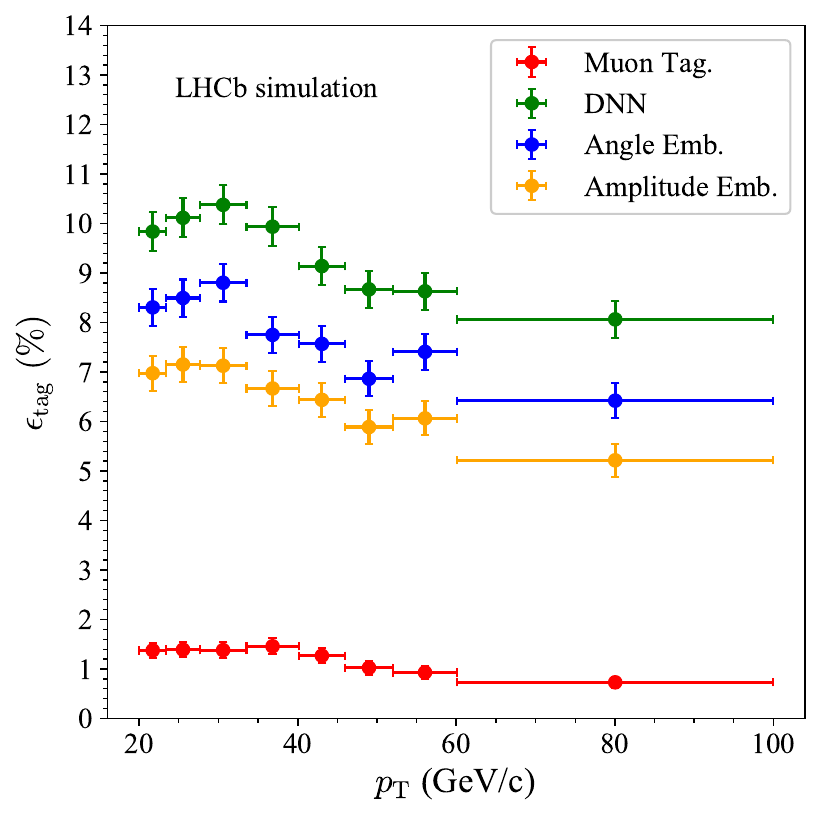}
    \caption{}
    \label{fig:tagpower_left}
    \end{subfigure}
    \hfill
    \begin{subfigure}[b]{0.49\textwidth}
    \centering
    \includegraphics[width=\textwidth]{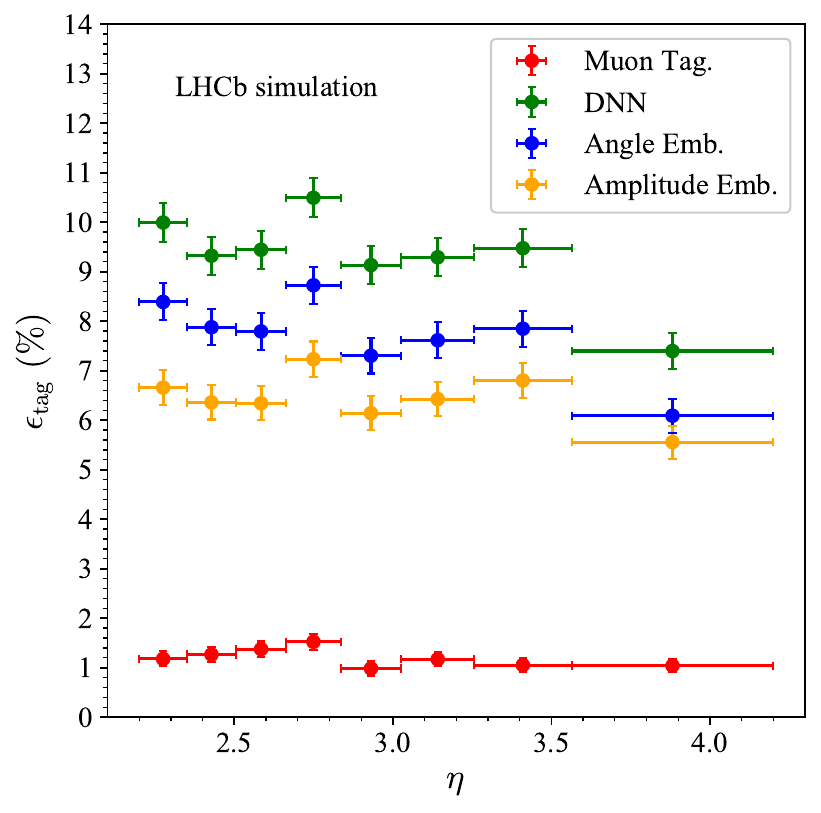}
    \caption{}
    \label{fig:tagpower_right}
    \end{subfigure}
    \caption{Tagging power $\epsilon_{\text{tag}}$ with respect to (a) jet $p_{\mathrm{T}}$  and (b) jet $\eta$  for the \emph{complete dataset}. The quantum algorithms perform slightly worse than the DNN, with the Angle Embedding circuit performing better than the Amplitude Embedding circuit.}
    \label{fig:tagpower}
\end{figure}

 The tagging power $\epsilon_{\text{tag}}$ for the DNN and the quantum classifiers on the \emph{complete dataset} as function of jet $p_{\mathrm{T}}$ and $\eta$, is shown in Fig.~\ref{fig:tagpower}. Also in this case, the usual dependence on the jet $p_{\mathrm{T}}$ is visible. As expected, the tagging power of the QML and DNN is higher in the \emph{complete dataset} with respect to the \emph{muon dataset}, since a larger number of features is used. As before, for QML and DNN the performance is above the muon tagging approach, given that these classifiers use the information coming from the jet substructure.
 Differently from the application to the \emph{muon dataset}, in the \emph{complete dataset} the QML algorithms perform slightly worse than the DNN, with slightly better performance for the Angle Embedding structure than the Amplitude Embedding. It can be deduced that the DNN makes a better use of the features when a larger number of them is used.

\subsection{Dependence of the results on number of training events and circuit depth}
\label{subsec:numbereventscircuitdepth}
\begin{figure}[tb]
    \centering
    \begin{subfigure}[b]{0.49\textwidth}
    \centering
    \includegraphics[width=\textwidth]{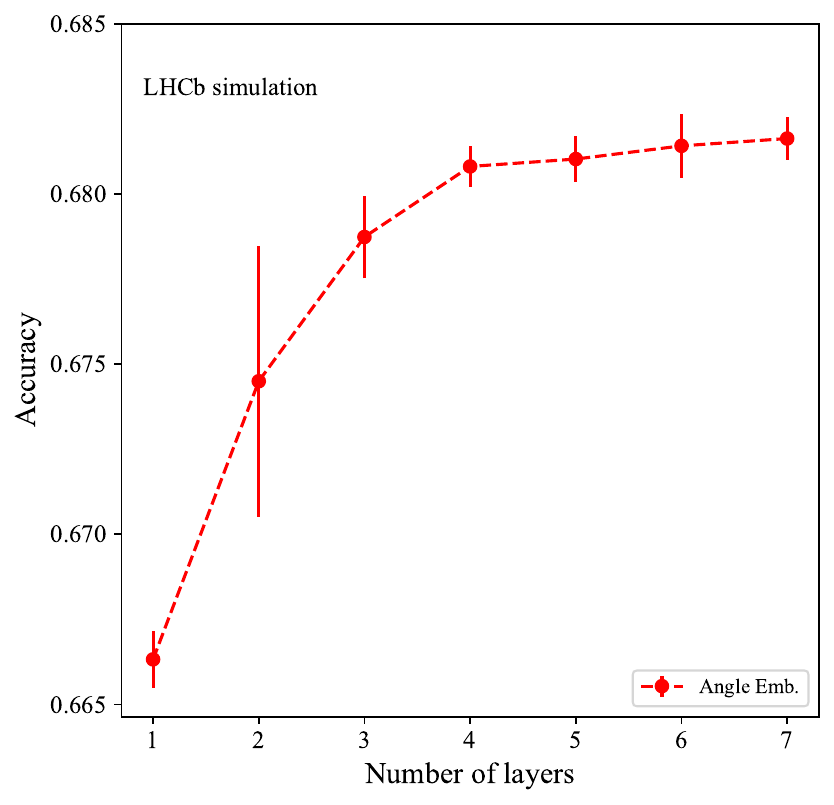}
    \caption{}
    \label{fig:acc_wrt_layers_left}
    \end{subfigure}
    \hfill
    \begin{subfigure}[b]{0.49\textwidth}
    \centering
    \includegraphics[width=\textwidth]{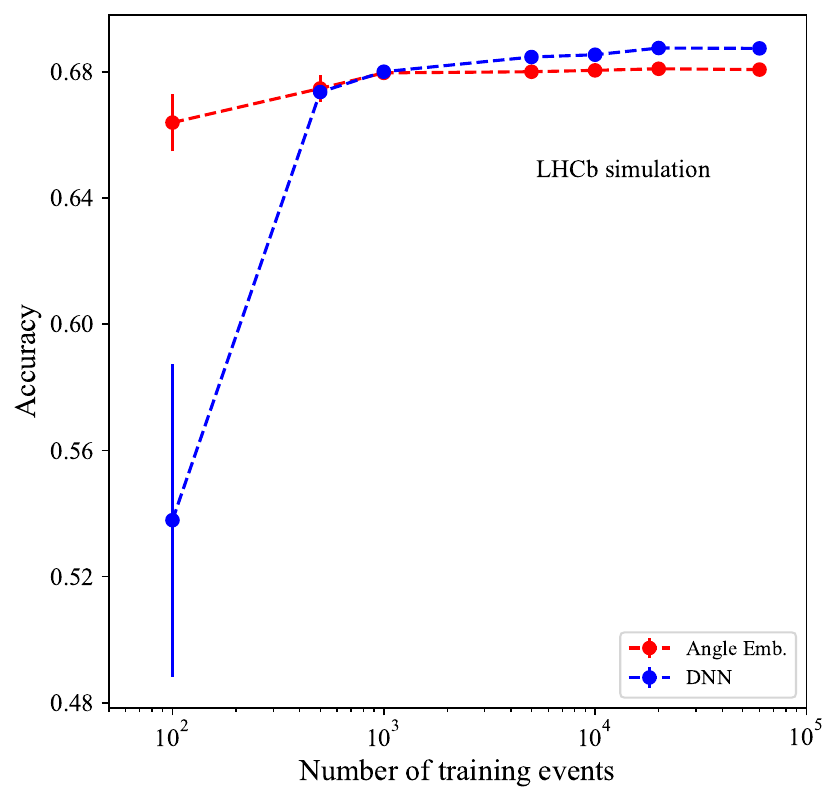}
    \caption{}
    \label{fig:acc_wrt_layers_right}
    \end{subfigure}
    \caption{(a) Accuracy of the Angle Embedding structure on the \emph{muon dataset} versus the number of layers. (b) Accuracy of the (red) Angle Embedding structure  and (blue) DNN on the \emph{muon dataset} versus the number of training events.}
    \label{fig:acc_wrt_layers}
\end{figure}

The dependence of the quantum algorithms performance on the number of training events and the circuit complexity has to be evaluated if near-term applications on quantum hardware are considered. These parameters have an impact on the execution times and therefore on the possibility to use it. The performance dependence on the number of training samples is an interesting parameter to compare QML and DNN methods, in order to assess the differences between the two approaches. Given the high computational efforts of simulating complex circuits with several qubits, only the \emph{muon dataset} is used.
For QML, the Angle Embedding structure is considered with different number of strongly entangled layers and different number of training events. The results are compared with the same DNN considered in the previous section; a comparison with more complex networks is described in App.~\ref{app:DNN}.
The metric used to quantify the goodness of the quantum classifier is the accuracy on a test subset of 40000 jets. The performance is calculated  averaging over 10 training rounds. In Fig.~\ref{fig:acc_wrt_layers} (a) the accuracy of the Angle Embedding circuit is shown as a function of the number of layers of the circuit. As expected, by increasing the depth of the circuit, and therefore its complexity, the accuracy increases. This behaviour stops at around 5 layers, where the accuracy is saturating and no further improvement is evident.
It is clear that, for a given number of features and training data, the Angle Embedding model does not profit of an arbitrarily large number of layers, therefore it is possible to keep a low number of layers, and subsequently a lower complexity of the circuit, to obtain the best performance. This would reduce also the computing time and resources needed for the simulation.

The accuracy as a function of the number of training events for the Angle Embedding circuit and the DNN is shown in Fig.~\ref{fig:acc_wrt_layers} (b). Increasing the number of training events the performance of the quantum algorithm is similar to the DNN, but when the number of training events decreases the quantum algorithm keeps very high performance, while the DNN is not able to perform a good classification.  
This means that, with respect to the DNN, the QML method reaches optimal performance with a lower number of events. Considering the fact that state-of-the-art ML algorithms require very large data sets to get meaningful performance, this unique feature of QML algorithms needs further investigation, which could lead to a better understanding on how these algorithms are using the input features.

\subsection{Time performance}
\begin{figure}[tb]
    \centering
    \includegraphics[width=0.6\textwidth]{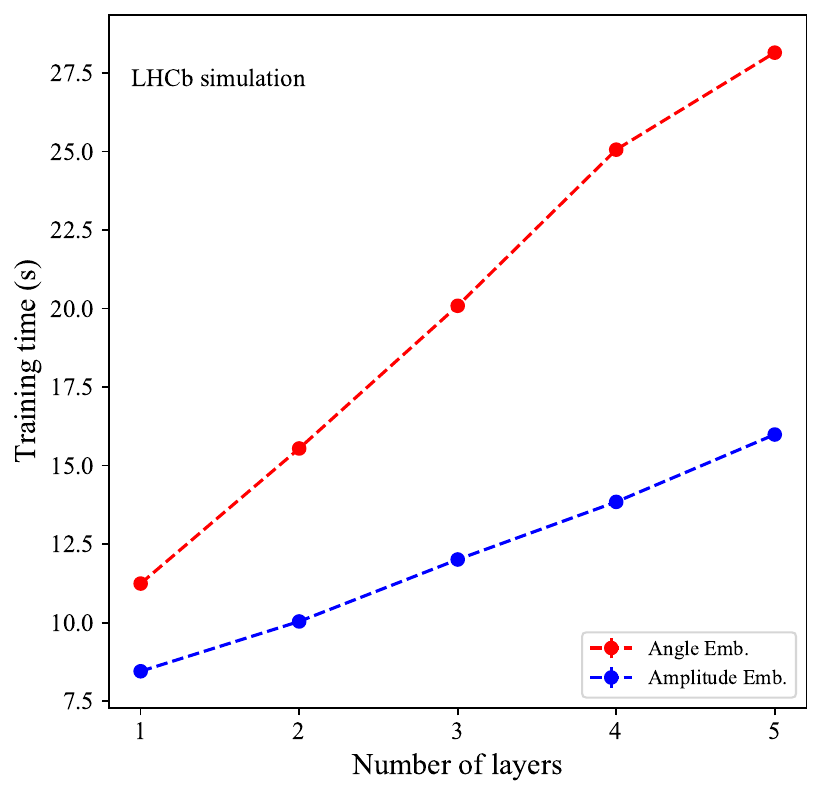}
    \caption{Training time for 100 epochs for the Angle Embedding (red) and Amplitude Embedding (blue) circuits on the \emph{muon dataset} with respect to the number of circuit layers.}
    \label{fig:time}
\end{figure}

Time performance is a fundamental figure of merit to understand the feasibility of simulating such quantum algorithms. Here, the time performance for the Angle Embedding and the Amplitude Embedding circuits is evaluated for the \emph{muon dataset}, as a function of the number of strongly entangling layers in the circuit structure. The quantum algorithms are trained using 4 NVIDIA Volta V100 GPUs, and the training time to train 60000 jets for 100 epochs is evaluated. Results are shown in Fig.~\ref{fig:time}.
Results show that it takes less time to train the Amplitude Embedding circuit than the Angle Embedding structure, and the training time increases as the number of layers, although with a greater rate for the Angle Embedding circuit: this may be expected since the complexity of the circuit increases with the number of layers, therefore it takes more time to simulate the quantum circuit.
\subsection{Noise models}
\label{subsec:noisemodels}
\begin{figure}[tb]
    \centering
    \begin{subfigure}[b]{0.49\textwidth}
    \centering
    \includegraphics[width=\textwidth]{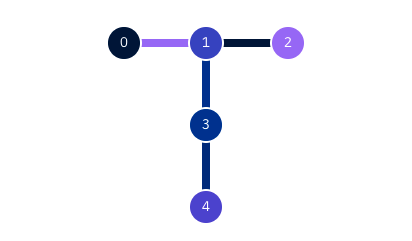}
    \caption{}
    \label{fig:circuits_noise_belem}
    \end{subfigure}
    \hfill
    \begin{subfigure}[b]{0.49\textwidth}
    \centering
    \includegraphics[width=\textwidth]{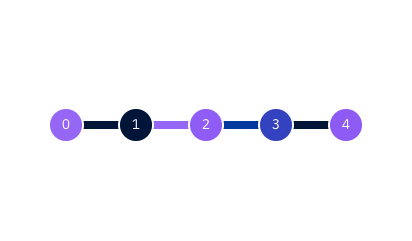}
    \caption{}
    \label{fig:circuits_noise_santiago}
    \end{subfigure}
    \caption{Qubit structure of \texttt{ibmq-belem} (a) and \texttt{ibmq-santiago} (b).}
    \label{fig:circuits_noise}
\end{figure}

In order assess the performance of the algorithms in quantum hardware it is important to understand the impact of noise on quantum circuits. Two kinds of noise can affect quantum algorithms:
\begin{itemize}
    \item coherent noise: it originates from unitary errors in the application of quantum gates. This lead to the construction of a different quantum state with respect to the desired one. A typical source of this kind of noise is non-ideal calibrations of the quantum hardware;
    \item incoherent noise: this noise results from the interaction between the quantum hardware and the environment. This noise gives quantum states that are not pure anymore and are described by mixed states, \emph{i.e.} probability distributions over different states.
\end{itemize}
The simulations of noise contribution taking into account both sources of noise in quantum circuit measurements have been performed using the \emph{pennylane-qiskit} plugin~\cite{ref:pennylane,ref:qiskit}. This plugin allows to simulate noise models coming from different real IBM quantum computers~\cite{ref:ibmq}, including state preparation and readout errors, and to  keep the Pennylane syntax.  The result is a simulation of a quantum algorithm on a real device structure. Four IBM quantum computers are considered: \texttt{ibmq-belem}, \texttt{ibmq-santiago}, \texttt{ibmq-jakarta} and \texttt{ibmq-toronto}, which have different numbers of qubits (respectively 5, 5, 7 and 27 qubits), different quantum volumes\footnote{The quantum volume is the maximum size of a quantum circuit that can be effectively implemented on a noisy intermediate-scale quantum device. In this paper, the definition from Ref.~\cite{ref:quantumvolume} is adopted.} (respectively 16, 32, 16 and 32) and different qubits structure, as shown in Fig.~\ref{fig:circuits_noise} for the \texttt{ibmq-santiago} and \texttt{ibmq-belem} which have the same number of qubits.

\begin{figure}[tb]
    \centering
    \includegraphics[width=0.6\textwidth]{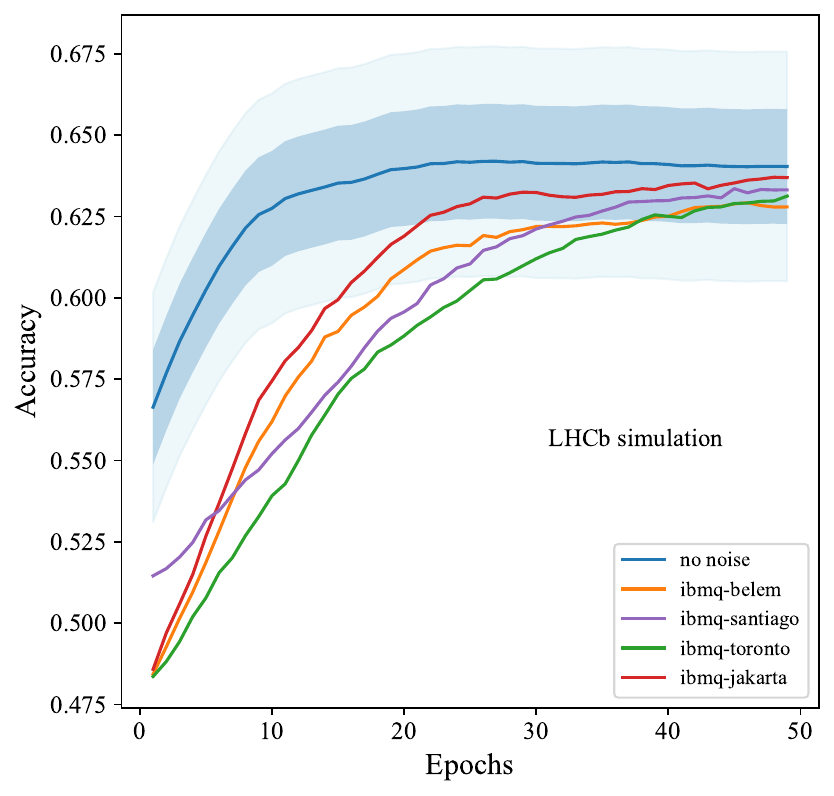}
    \caption{Validation accuracy for noise models as a function of the number of epochs. Blue (light-blue) band represent $1\sigma$ ($2\sigma$) uncertainty bounds for the noiseless model.}
    \label{fig:acc_noise}
\end{figure}

Studies are performed on the Angle Embedding circuit structure with three strongly entangled layers.  A small subset of the \emph{muon dataset} is used  because  simulating circuits including noise contribution is more time and computationally consuming; on the other hand, with a low number of events the quantum algorithm performance is sufficiently high, as shown in Fig.\ref{fig:acc_wrt_layers}.
In this way, a subset of 1000 jets of the \emph{muon dataset} is selected for training while validation is performed on a subset of 10000 jets.  For each noise model the training is performed for 50 epochs using ADAM with a learning rate $\xi=0.01$ and batch size of 10 jets. The results are averaged over five rounds of training, using five independent training subsets. The relevant figure of merit to assess noise models performance is the accuracy on the validation test.
The results are shown in Fig.~\ref{fig:acc_noise} and summarised in Tab.~\ref{table}.  Models including noise need more epochs to reach convergence, but in the end the results are consistent with those of noiseless simulations within error. Such a result demonstrates  that the proposed circuit model for the \emph{muon dataset} is robust to noise. 

\begin{table}[tb]
\caption{Accuracy for noisy circuits obtained with the \emph{muon dataset}.}
\label{table}
\centering
\begin{tabular}{|c|c|}
\hline
Noise model & accuracy \\
\hline
\hline
no noise & $0.640 \pm 0.017$\\
\hline
\texttt{ibmq-belem} & $0.629 \pm 0.047$\\
\texttt{ibmq-santiago} & $ 0.633\pm 0.038$\\
\texttt{ibmq-jakarta} & $0.637 \pm 0.042$\\
\texttt{ibmq-toronto} & $ 0.631 \pm 0.044$ \\
\hline
\end{tabular}
\end{table}

\section{Conclusions}
\label{sec:conclusions}
The first application of QML algorithms to identify the charge of the $b$ hadron produced in the jet hadronization, at the LHCb experiment has been presented. The results using the \emph{muon dataset} show that the Angle Embedding structure almost reaches the performance of the DNN while being better than the muon tagging approach. The amplitude encoding structure presents no evident improvement with respect to the muon tagging method.
When the \emph{complete dataset} is used, the best quantum algorithm performance is given by Angle Embedding structure.
The study of the performance dependence on the circuit depth has shown that the number of layers is a parameter to be optimised. More layers, which means more complex structures, do not necessarily result in improved performance, since saturation is reached.
The impact of the noise on the results appears to be negligible, suggesting that  the proposed circuits could be implemented on a quantum hardware if available.
QML algorithms achieve performance consistent with classical methods like the DNN with low-complexity circuits and a smaller number of training events. This could have important implications for LHC experiments where often the training phase is the most expensive in terms of resources.
However, when a large number of features is employed, the DNN performs better than QML algorithms. Here huge improvements are expected when the hardware will be available. In fact, the comparison of QML models  to classical kernel methods~\cite{ref:quantumkernel} shows that QML models achieve classification tasks separating data in Hilbert spaces whose dimension scales exponentially with the number of qubits involved. As this number increases, quantum simulations on classical computers become unfeasible.

The  full exploitation of QML in high energy physics experiments at colliders, just began. As for any brand new tool, unexpected applications may manifest themselves while using. 
For example, results obtained involving tensor network methods~\cite{ref:ttnhep,ref:ttn} have shown that quantum algorithms can allow to study correlations among the features. This is done by measuring the entanglement correlations between qubits of an optimised multi-qubit system. Given that a quantum circuit is a unitary matrix, the same approach applied to tensor networks can be applied in QML by easily inverting the unitary matrix representing the circuit. This new and yet not investigated approach could allow to extract information on jet constituents correlations starting from the measurement of the entanglement between the qubits of a trained VQCs and possibly help in improving the classification performance.

\section*{Acknowledgements}
The authors would like to thank the LHCb Data Processing \& Analysis (DPA) project colleagues for supporting this publication and reviewing the work. 
The authors acknowledge support from INFN (Italy) and from the Department of Physics and Astronomy of the University of Padova. The authors would also like to thank CERN Openlab for useful and fruitful discussions.
The authors thank the CINECA Supercomputing Center for awarding access to Marconi100 cluster based at CINECA (Bologna, Italy).

\appendix

\section{Deep Neural Network model and further comparisons}
\label{app:DNN}
\begin{figure}[tb]
    \centering
    \includegraphics[width=0.65\textwidth]{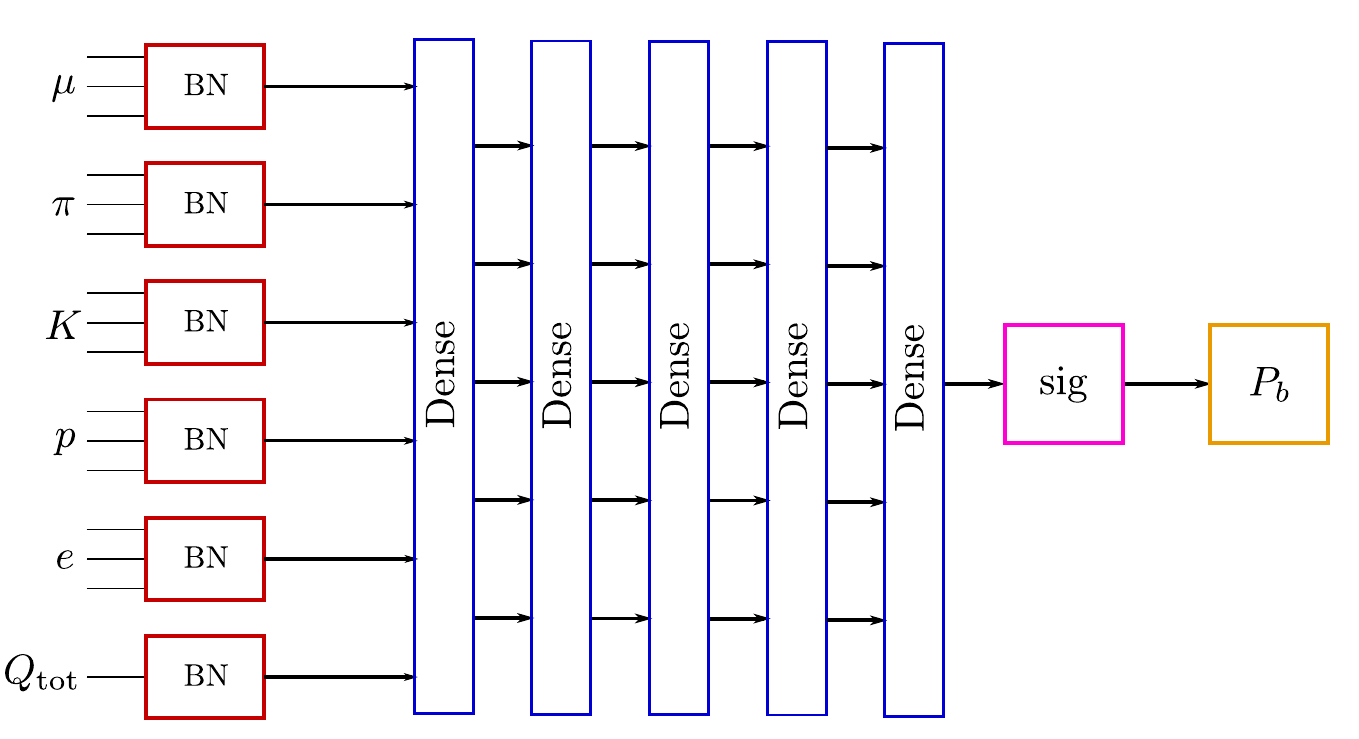}
    \caption{DNN structure.}
    \label{fig:DNN}
\end{figure}

\begin{figure}[tb]
    \centering
    \begin{subfigure}[b]{0.495\textwidth}
    \centering
    \includegraphics[width=\textwidth]{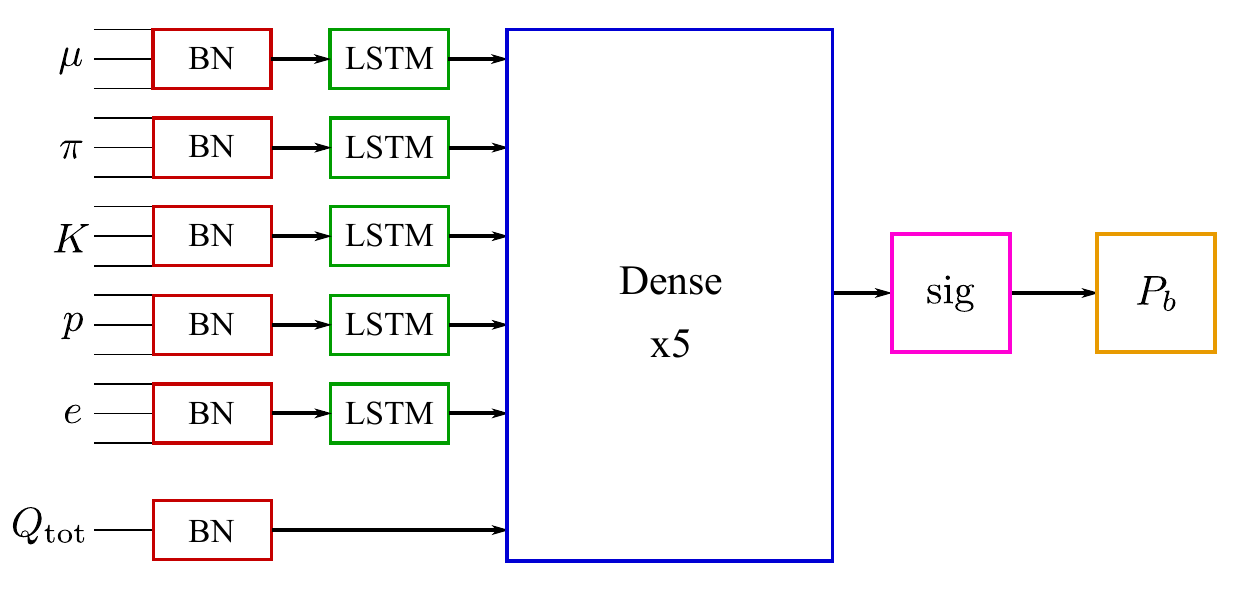}
    \caption{}
    \label{fig:LSTM}
    \end{subfigure}
    \hfill
    \begin{subfigure}[b]{0.495\textwidth}
    \centering
    \includegraphics[width=\textwidth]{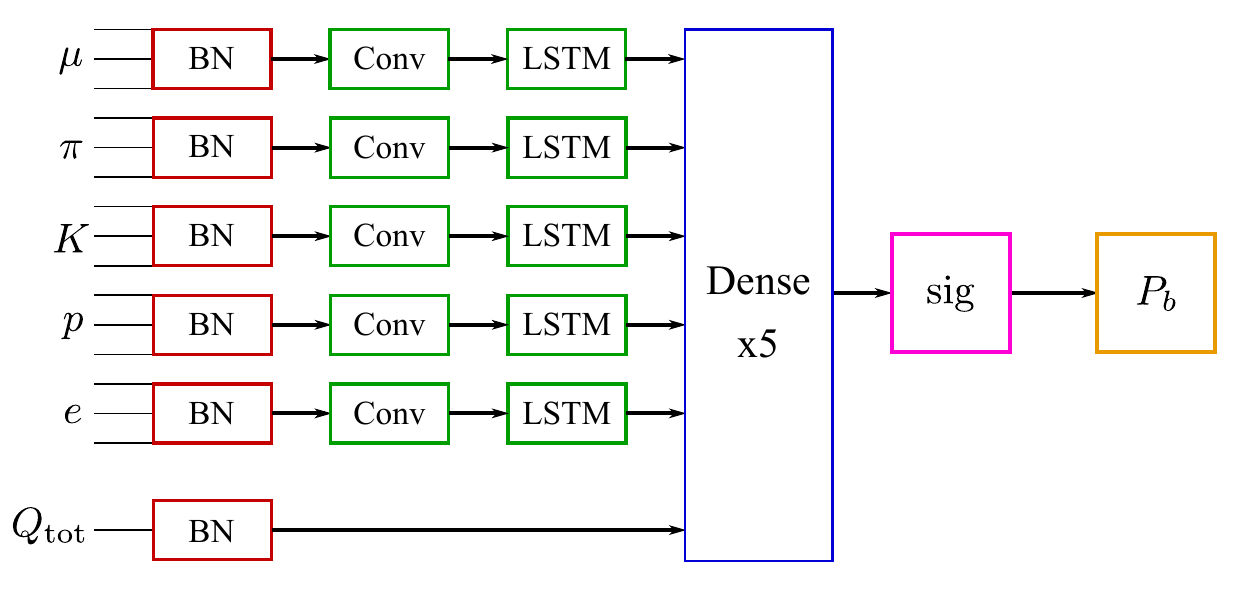}
    \caption{}
    \label{fig:LSTM+CONV}
    \end{subfigure}
    \caption{DNN structure for (a) the ``LSTM" model  and for (b) the ``LSTM+CONV" model .}
    \label{fig:LSTM/LSTM+CONV}
\end{figure}

\begin{figure}[tb]
    \centering
    \begin{subfigure}[b]{0.49\textwidth}
    \centering
    \includegraphics[width=\textwidth]{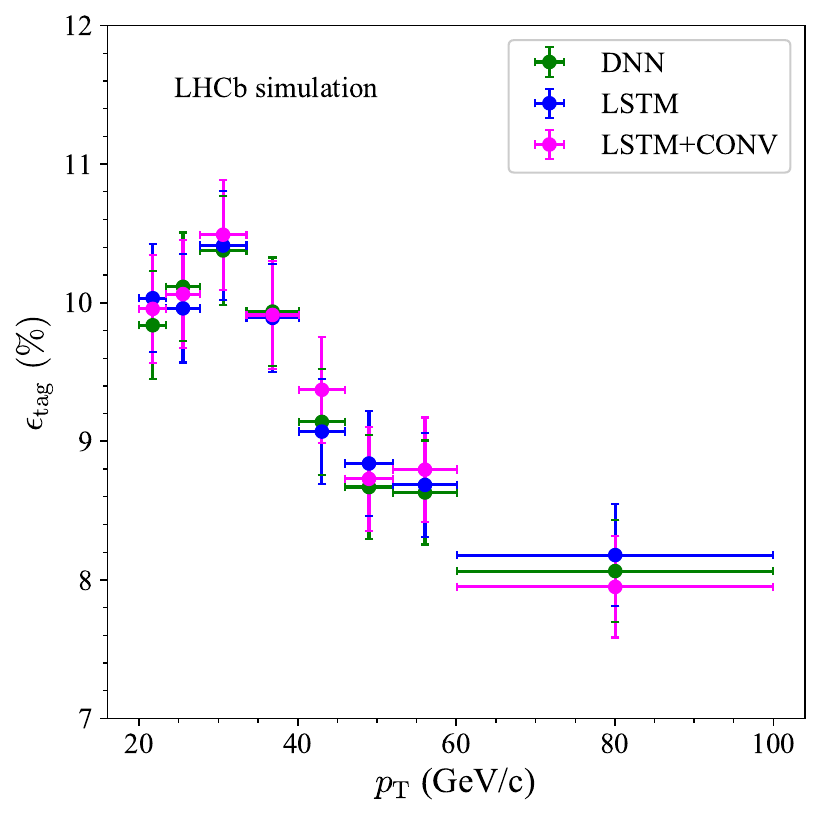}
    \caption{}
    \label{fig:LSTM}
    \end{subfigure}
    \hfill
    \begin{subfigure}[b]{0.49\textwidth}
    \centering
    \includegraphics[width=\textwidth]{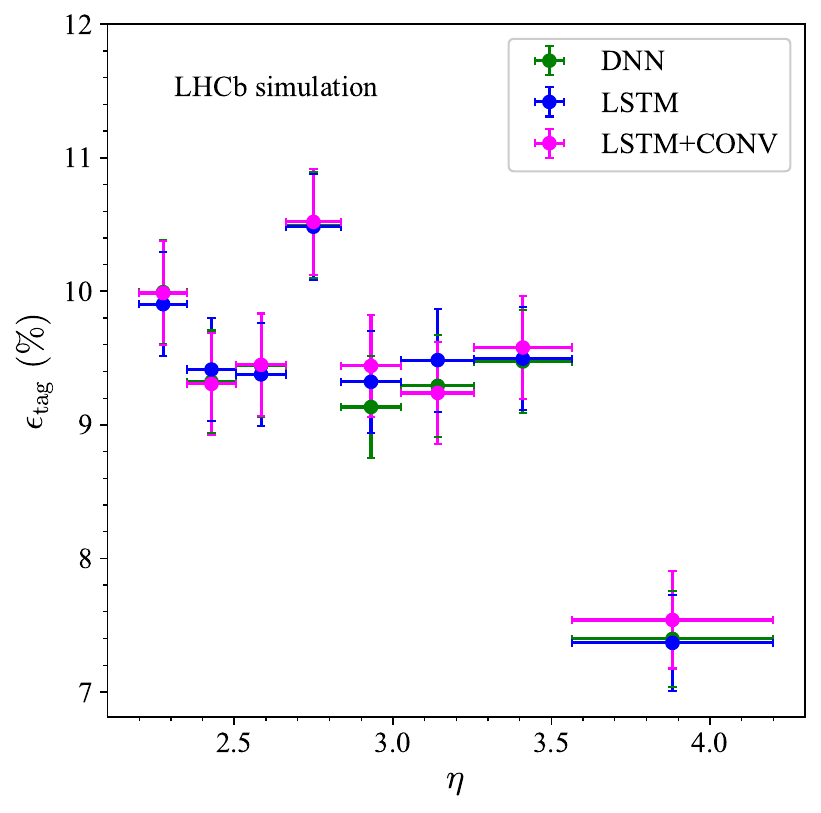}
    \caption{}
    \label{fig:LSTM+CONV}
    \end{subfigure}
    \caption{Tagging power $\epsilon_{\text{tag}}$ with respect to (a) jet $p_\mathrm{T}$  and (b) jet $\eta$  for the complete dataset applied to different DNN models.}
    \label{fig:tagpower_LSTM/CONV}
\end{figure}

 The DNN is built following a classical feed-forward structure shown in Fig.~\ref{fig:DNN}: it starts with a Batch Normalisation Layer\cite{ref:batchnorm} and it applies several Dense layers, each one followed by a Dropout\cite{ref:dropout} layer. Depending on the number of input features the input vector for the DNN changes (4 variables for the muon dataset and 16 for the complete one). For the hidden layers the ReLu\footnote{ReLu is the rectifier activation function, defined as $\mathrm{ReLu}(x) = \max(0,x)$} activation is used while a sigmoid\footnote{The sigmoid activation function is defined as $\mathrm{sig}(x) = [1 + \exp{(-x)}]^{-1}$} function is applied to the output node. The network is trained using the ADAM optimiser\cite{ref:adam}. The model is trained for 250 epochs with an early stopping of 25 epochs on the test loss function and a learning rate of 0.0001. The hyperparameters of the DNN (such as depth, number of nodes per layer, dropout rate, normalization moment, and kernel initialization) are optimised using the \emph{hyperopt} package~\cite{ref:hyperopt}, maximising the accuracy in the test dataset by exploring different parameter spaces. Given the classification task, we take cross entropy as loss function. A scheme of the DNN model is shown in Fig.~\ref{fig:DNN}.
Different DNN structures have also been considered, in order to have a fair comparison between quantum and classical classifiers. Usual state-of-the-art networks, such as CMS DeepJet algorithm~\cite{ref:deepjet}, make use of LSTM~\cite{ref:lstm} and convolutional layers, resulting in an improvement of performance. RNN structures are also under study within the LHCb experiment as potential flavour tagging algorithms. Therefore the following DNN models are studied:
\begin{itemize}
    \item ``LSTM" model: starting from the DNN structure we apply a LSTM layer to the particle features (therefore excluding the "global" variable of the total jet charge) before the first Dense layer. The Dense structure is identical to the DNN structure.
    \item ``LSTM+Conv" model: from the LSTM model we firstly apply a convolutional layer of the particle features (as before not applied to the total jet charge).
\end{itemize}
The two models are shown in Fig.~\ref{fig:LSTM/LSTM+CONV}.
The DNN performance in terms of tagging power is compared to the LSTM and LSTM+CONV models. Results for tagging power as function of the jet $p_\mathrm{T}$ and $\eta$ are shown in Fig.~\ref{fig:tagpower_LSTM/CONV}, where results for different models are comparable within the error and therefore allow us to consider just the DNN model for the comparison with quantum algorithms.

\section{Unoptimised tagging power distributions}
\label{app:noOptTagPower}
\begin{figure}[tb]
    \centering
    \begin{subfigure}[b]{0.49\textwidth}
    \centering
    \includegraphics[width=\textwidth]{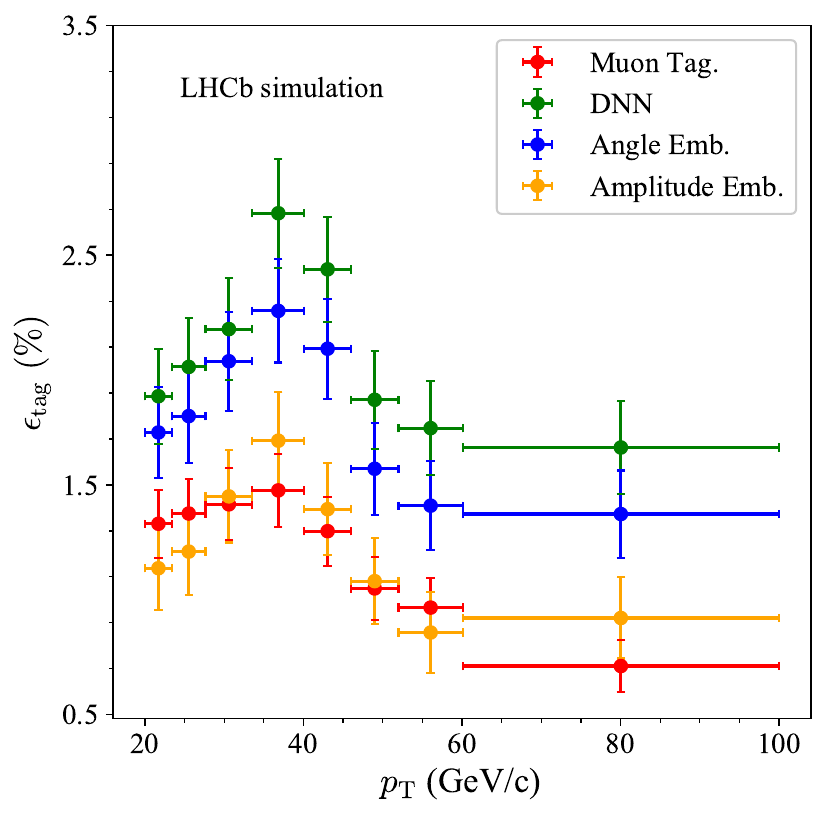}
    \caption{}
    \label{fig:tagpower_mu_noOpt_left}
    \end{subfigure}
    \hfill
    \begin{subfigure}[b]{0.49\textwidth}
    \centering
    \includegraphics[width=\textwidth]{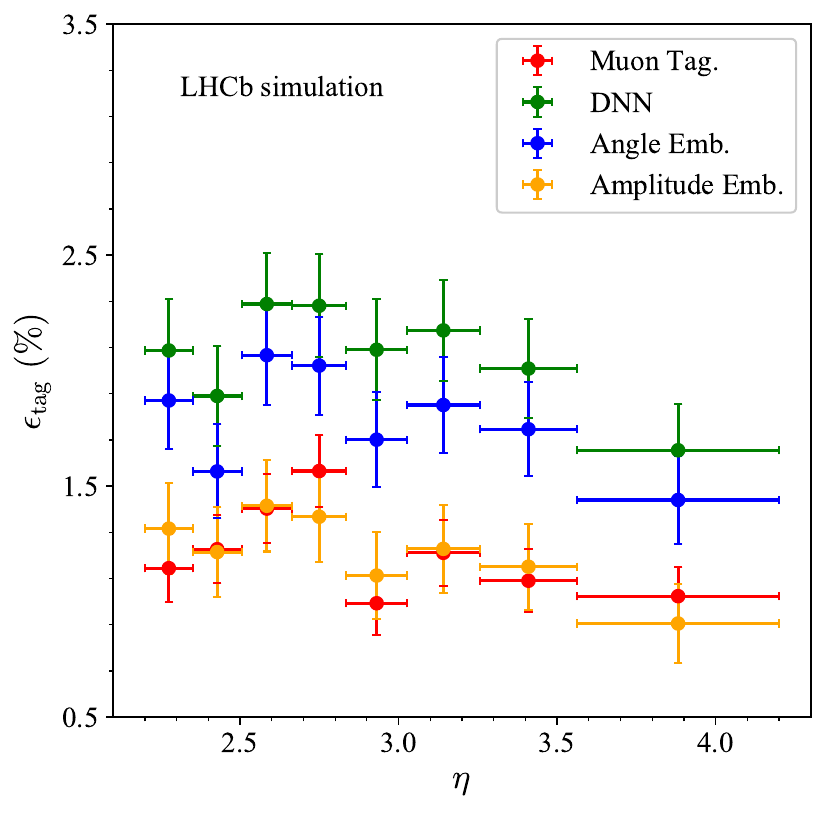}
    \caption{}
    \label{fig:tagpower_mu_noOpt_right}
    \end{subfigure}
     
    \caption{Unoptimised tagging power $\epsilon_{\text{tag}}$ with respect to (a) jet $p_\mathrm{T}$  and (b) jet $\eta$  for the muon dataset. The angle embedding circuit and the DNN show similar performance.}
    \label{fig:tagpower_mu_noOpt}
\end{figure}

\begin{figure}[tb]
    \centering
    \begin{subfigure}[b]{0.49\textwidth}
    \centering
    \includegraphics[width=\textwidth]{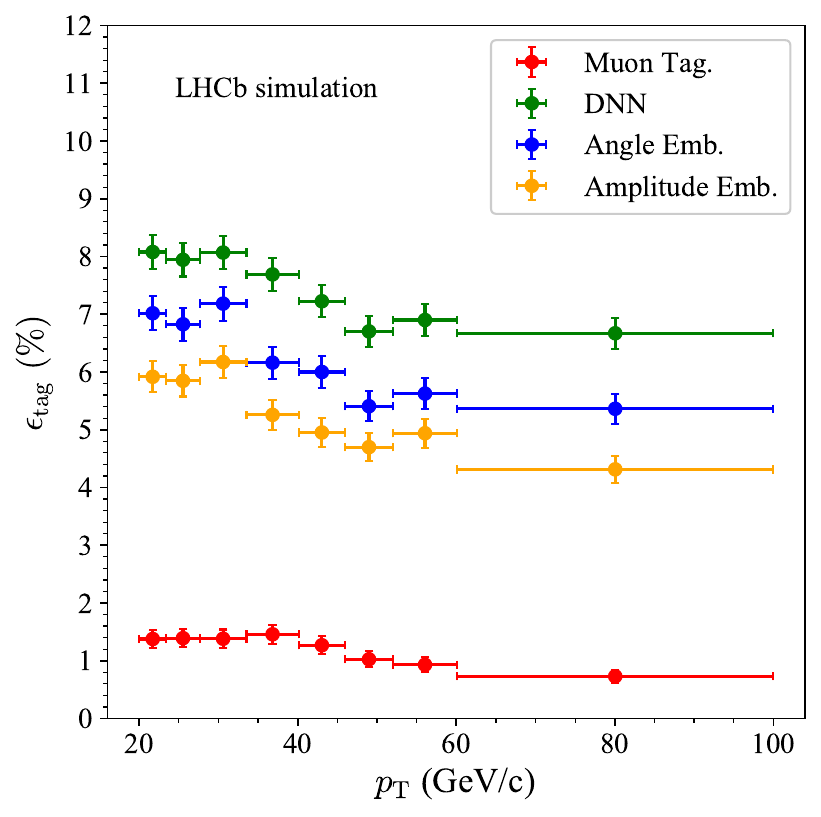}
    \caption{}
    \label{fig:tagpower_noOpt_left}
    \end{subfigure}
    \hfill
    \begin{subfigure}[b]{0.49\textwidth}
    \centering
    \includegraphics[width=\textwidth]{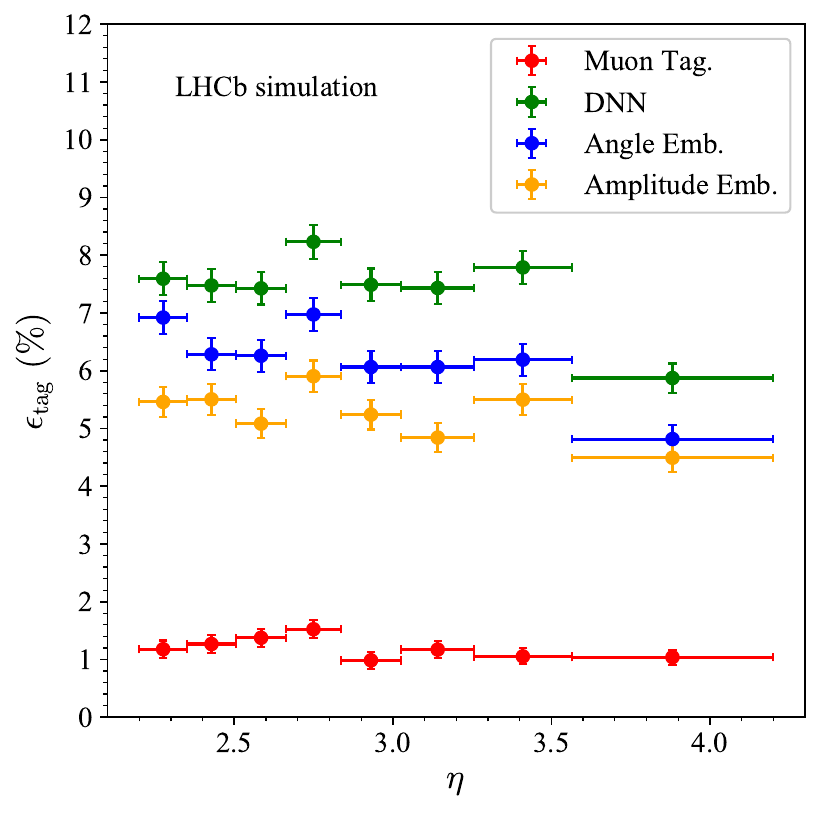}
    \caption{}
    \label{fig:tagpower_noOpt_right}
    \end{subfigure}
     
    \caption{Unoptimised tagging power $\epsilon_{\text{tag}}$ with respect to (a) jet $p_\mathrm{T}$  and (b) jet $\eta$  for the complete dataset. The angle embedding circuit and the DNN show similar performance.}
    \label{fig:tagpower_noOpt}
\end{figure}

In this section the distributions for the unoptimised tagging power are shown, both for the muon and the complete datasets. This is done in order to check if there are performance biases when cutting on the probability distributions to maximise the tagging power. In Figs.~\ref{fig:tagpower_mu_noOpt} and \ref{fig:tagpower_noOpt} the unoptimised tagging power distributions as function of jet $p_\mathrm{T}$ and $\eta$ respectively for the muon and the complete datasets are shown. No evident biases are present and the same considerations done for the optimised distributions are valid.

\clearpage

\addcontentsline{toc}{section}{References}
\bibliographystyle{LHCb}
\bibliography{references}

\end{document}